\documentclass[twocolumn, twocolappendix]{aastex701}

\usepackage{amsmath}
\usepackage{graphicx}
\usepackage{txfonts}                        
\usepackage{placeins}          
\usepackage{orcidlink}
\usepackage{booktabs}
\usepackage{twoopt}
\usepackage{enumerate}
\usepackage{url}
                                
\usepackage{natbib}
\definecolor{cobalt}{rgb}{0.06, 0.2, 0.65}
\hypersetup{
  colorlinks,
  citecolor=blue,
  linkcolor=blue,
  urlcolor=cobalt,
  pdftitle={A two-stage settling of the Milky Way disk revealed by precise ChronoGal ages}, pdfauthor={David Mirabal}
}

\defcitealias{Gallart2024}{Paper~I}
\defcitealias{emma2025}{Paper~II}
\defcitealias{yllari2025}{Paper~III}
\defcitealias{tomas2025}{Paper~IV}

\defcitealias{gaiadr3}{Gaia Collaboration et al. 2023}

\begin{document}

%%%%%%%%%%%%%%%%%%%%%%%%%%%%%%%%%%%%%%%%

\title{A two-stage settling of the Milky Way disk revealed by precise ChronoGal ages}

%%%%%%%%%%%%%%%%%%%%%%%%%%%%%%%%%%%%%%%%

\author[orcid=0009-0005-1814-0671]{David Mirabal}
\affiliation{Instituto de Astrofísica de Canarias, E-38206 La Laguna, Tenerife, Spain}
\affiliation{Departamento de Astrofísica, Universidad de La Laguna, E-38205 La Laguna, Tenerife, Spain}
\email[show]{david.mirabal@iac.es}  

\author[orcid=0000-0001-6728-806X]{Carme Gallart} 
\affiliation{Instituto de Astrofísica de Canarias, E-38206 La Laguna, Tenerife, Spain}
\affiliation{Departamento de Astrofísica, Universidad de La Laguna, E-38205 La Laguna, Tenerife, Spain}
\email{carme.gallart@iac.es}

\author[orcid=0000-0001-6984-4795]{Tomás Ruiz-Lara}
\affiliation{Departamento de Física Teórica y del Cosmos, Universidad de Granada, Campus Fuente Nueva, Edificio Mecenas, E-18071 Granada, Spain}
\affiliation{Instituto Carlos I de Física Teórica y Computacional, Universidad de Granada, E-18071 Granada, Spain}
\email{tomasruizlara@gmail.com}

\author[orcid=0000-0001-8994-0936]{Emma Fernández-Alvar}
\affiliation{Instituto de Astrofísica de Canarias, E-38206 La Laguna, Tenerife, Spain}
\affiliation{Departamento de Astrofísica, Universidad de La Laguna, E-38205 La Laguna, Tenerife, Spain}
\email{emma.fernandez@iac.es}

\author[orcid=0000-0001-9209-7599]{Anna B. A. Queiroz}
\affiliation{Instituto de Astrofísica de Canarias, E-38206 La Laguna, Tenerife, Spain}
\affiliation{Departamento de Astrofísica, Universidad de La Laguna, E-38205 La Laguna, Tenerife, Spain}
\email{anna.barbara@iac.es}

\author[orcid=0009-0000-2002-2416]{Guillem Aznar-Menargues}
\affiliation{Instituto de Astrofísica de Canarias, E-38206 La Laguna, Tenerife, Spain}
\affiliation{Departamento de Astrofísica, Universidad de La Laguna, E-38205 La Laguna, Tenerife, Spain}
\email{guillem.aznar@iac.es}

\author[orcid=0000-0002-0534-4115]{Chris B. A. Brook}
\affiliation{Instituto de Astrofísica de Canarias, E-38206 La Laguna, Tenerife, Spain}
\affiliation{Departamento de Astrofísica, Universidad de La Laguna, E-38205 La Laguna, Tenerife, Spain}
\email{chbrook@ull.edu.es}

\author[orcid=0000-0001-5870-3735]{Santi Cassisi}
\affiliation{INAF -- Astronomical Observatory of Abruzzo, Via M. Maggini, sn, 64100 Teramo, Italy}
\affiliation{INFN, Sezione di Pisa, Largo Pontecorvo 3, 56127 Pisa, Italy}
\email{santi.cassisi@inaf.it}

\author[orcid=0009-0001-3430-9140]{Yllari K. González-Koda}
\affiliation{Departamento de Física Teórica y del Cosmos, Universidad de Granada, Campus Fuente Nueva, Edificio Mecenas, E-18071 Granada, Spain}
\email{yllari@ugr.es}

\author[orcid=0000-0003-3862-5076]{Julio F. Navarro}
\affiliation{Physics Department, University of Victoria, Victoria, BC V8P1A1, Canada}
\affiliation{KU-KIST Green School, Korea University}
\email{jfn@uvic.ca}

\author[orcid=0000-0002-7432-8709]{Pedro A. Palicio}
\affiliation{Instituto de Astrofísica de Canarias, E-38206 La Laguna, Tenerife, Spain}
\affiliation{Departamento de Astrofísica, Universidad de La Laguna, E-38205 La Laguna, Tenerife, Spain}
\email{pedro.alonso.palicio@iac.es}

\author{Francisco Surot}
\affiliation{Instituto de Astrofísica de Canarias, E-38206 La Laguna, Tenerife, Spain}
\affiliation{Departamento de Astrofísica, Universidad de La Laguna, E-38205 La Laguna, Tenerife, Spain}
\email{frsurot@uc.cl}

\begin{abstract}

The spatial distribution of Galactic disk stars of different ages and metallicities encodes its dynamical evolution history and formation pathways. The lack of precise ages for large, representative samples of stars has traditionally been a drawback for securely differentiating among formation scenarios proposed to explain the vertical properties of the Milky Way disk (e.g. upside-down formation {\it versus} secular dynamical heating). In this Letter, we use CMD-fitting of \textit{Gaia} data to derive very precise age-metallicity distributions for complete stellar samples in the solar neighborhood. We identify quasi-mono-age-metallicity stellar populations with an age resolution better than $5\%$ ($\simeq$ 0.5 Gyr at old age), and study their absolute vertical density profile. We observe a clear correlation between scale height and stellar age, with younger populations consistently more confined to the plane. We find evidence that the scale height decrease rate changed drastically $10.85^{+0.13}_{-0.12}$ Gyr ago: at older ages, the evolution is very steep, with $h_Z$ dropping from over 1.26 kpc to $\sim0.5$ kpc in merely 2 Gyr; subsequently, the thinning is very gradual, with $ h_Z \sim0.1$ kpc for the youngest populations. The extremely rapid early vertical settling of the disk points to a fundamentally \textit{upside-down} formation process at early times, when the thick disk component is basically formed. We thus strongly support an evolutionary scenario defined by two distinct temporal phases: an initial, highly turbulent epoch that gave birth to the thick disk, and a subsequent more quiescent period of secular evolution and gradual gas settling which formed the thin disk.

\end{abstract}

%%%%%%%%%%%%%%%%%%%%%%%%%%%%%%%%%%%%%%%%%%%%%%%%%%%%%%%%%%%%%%


%%%%%%%%%%%%%%%%%%%%%%%%%%%%%%%%%%%%%%%%%%%%%%%%%%%%%%%%%%%%%%

\section{Introduction}

The Milky Way (MW) disk is generally considered the superposition of two main components: a geometrically thin, dynamically cold and younger disk, versus a thicker, kinematically hotter and older counterpart \citep[e.g.,][]{Gilmore1983,  Fuhrmann1998, Bensby2003, Bensby2007, Bensby2014, Reddy2003, Soubiran2003, Juric2008, Adybekyan2011, Haywood2013}. The stars in these structures act as a fossil record that encodes the complex assembly history of our Galaxy. A central challenge in Galactic archaeology is to map the spatial distribution and temporal evolution of stellar populations within the disk, which should provide fundamental information for unraveling the physical processes governing its assembly.

 A number of studies have investigated the variation of the vertical scale height ($h_Z$) of stellar populations with metallicity and/or $\alpha$ abundances~\citep[the so-called mono-abundance populations -MAP-, e.g.,][]{Bovy2012thick, Bovy2012spatial, Bovy2016}, or also considering age estimates  \citep[e.g.,][]{Haywood2013, Mackereth2017, Imig2025}. These studies agree that $h_Z$ increases with stellar age \citep[e.g.][]{miglio2013, casagrande2016, Anders2023}, although \citet{Bovy2012thick} have questioned the mere existence of truly distinct thick and thin disk components. Similarly, the physical mechanisms driving the age-$h_Z$ relation are still debated. One paradigm is \textit{secular dynamical heating}, wherein stars are continuously born from a cold, thin gas layer and their orbits are progressively heated over time through gravitational scattering by non-axisymmetric structures such as giant molecular clouds, spiral arms, and the galactic bar \citep[e.g.,][]{Spitzer1951, Spitzer1953, Wielen1977, Nordstrom2004, Minchev2006}. An alternative (or complementary) paradigm is the \textit{upside-down} formation scenario. In this model, the early Galactic gas disk was highly turbulent and geometrically thick due to factors such as copious accretion, mergers, and intense feedback. Consequently, older stars were born inherently `hot', and the gas gradually cooled and settled over time to form the thin disk \citep[e.g.,][]{Brook2012, Bird2013, Haywood2013, Grand2016, Ma2017, Navarro2018, Gurvich23}.

Observational studies have so far not been able to provide constraints precise enough to discern between the above evolutionary scenarios. For example, based on a MW-like galaxy from a cosmological simulation, \citet{Ma2017} demonstrated that MAPs do not resolve the detailed structural evolution due to the large mixing of stellar ages within abundance bins. \citet{Minchev2017} reached a comparable conclusion using observational MW data. Similarly, most age estimates are not sufficiently precise to resolve trends in relatively narrow age bins. This is particularly true for the early MW evolution, for which precise ages are especially needed given the short timescales involved in the formation of a possibly distinct thick disk. Addressing this requires not only precise ages but also complete samples. This completeness is fundamental to accurately derive vertical scale heights, allowing us to disentangle early structural trends unaffected by the selection biases.

In the context of the ChronoGal project, to overcome this limitation and achieve highly precise ages of stellar populations ($\Delta\text{age}/\text{age}<0.05$), we employ Color-Magnitude Diagram (CMD) fitting applied to \textit{Gaia} data ($\textit{CMDft.Gaia}$) to derive star formation histories (SFHs) and their corresponding stellar age and metallicity distributions (AMDs). \citet[hereafter \citetalias{Gallart2024}]{Gallart2024} details the methodology and validation tests of this CMD fitting approach. $\textit{CMDft.Gaia}$ has proven successful in a variety of contexts (\citeauthor{emma2025} \citeyear{emma2025}, hereafter \citetalias{emma2025}, \citeauthor{yllari2025} \citeyear{yllari2025}, \citeauthor{tomas2025} \citeyear{tomas2025}). For further information, refer to \citetalias{Gallart2024}.

In this work, we perform a clustering analysis of the derived AMDs to identify quasi-mono-age-metallicity populations (qMAMs). This definition improves upon the traditional MAPs by adding the stellar age information, and allows us to identify stellar populations finely separated in age and metallicity, which most likely share a common  physical origin. We then demonstrate that qMAMs follow two distinct age-$h_Z$ relations, with a steeper slope before $\simeq$ 10.8 Gyr ago. For the first time, this clearly indicates the existence of two sets of physical mechanisms driving the evolution of the disk thickness.

This letter is organized as follows. In section~\ref{section:data}, we describe the data and the selection of the disk samples. Section~\ref{section:methodology} outlines the methodology, including CMD fitting, identification of qMAM populations, corrections to obtain the absolute number of stars, and modeling of vertical profiles. In Section~\ref{section:results}, we present the derived age-$h_Z$ relations. Finally, Section~\ref{section:discussion} discusses the physical implications of our findings, and Section~\ref{section:conclusions} summarizes our conclusions.

%%%%%%%%%%%%%%%%%%%%%%%%%%%%%%%%%%%%%%%%%%%%%%%%%%%%%%%%%%%%%%

\section{Data}\label{section:data}

We use data from \textit{Gaia} DR3 \citepalias{gaiadr3}. We applied the recommended photometric quality cuts as a function of color \citep[see e.g.][]{ruiz-lara2022} and corrected for reddening using the \cite{l22} extinction map. We also restricted our data to stars with low reddening ($A_\text{G} < 0.5$), and imposed \texttt{parallax\_over\_error~$~>~5$} after correcting for zero point parallax offsets. This data treatment has been standard across the papers of this collaboration \citepalias[e.g.][]{emma2025}.

\citetalias{emma2025} demonstrated that a classic kinematic selection of thick and thin disk samples resulted in a very clean separation of stellar populations older and younger than 10 Gyr. This separation helped to accurately characterize the older stellar populations: as they are progressively less separated in the CMD, they are inherently more challenging to recover using CMD fitting techniques, particularly when young stars dominate the CMD. Therefore, to accurately reconstruct the older populations, we isolated them from the younger stars in the CMD by defining kinematically cold and hot subsets, following a similar approach to that of \citet{Bensby2003}. The details of this selection are shown in Appendix~\ref{appendix:selection}. We restricted our analysis to $|Z| < 1$~kpc after verifying that, within this volume, the \textit{Gaia} DR3 RVS sample is complete for magnitudes brighter than the old main sequence turn-off. Subsequently, we defined two distinct cylindrical subsets:
\begin{enumerate}
    \item \texttt{cold}: A kinematically selected cold disk using the selection in Appendix~\ref{appendix:selection}, which considers the Galactic velocity components only. We restricted it to a small heliocentric radius ($R <250$~pc) to limit as much as possible the reddening effects, which are important close to the plane where most \texttt{cold} disk stars lie.
       
    \item \texttt{hot}: Applying the same kinematic selection method used in the previous sample, we defined a hot disk centered on the Sun with $R<750$~pc. Crucially, this dynamical selection ensures that the subset consists of co-rotating stars, keeping contamination from the Galactic halo negligible (below 10\% at least, see Appendix~\ref{appendix:selection}). The justification for the increased radius to select \texttt{hot} disk stars is twofold: i) as they are not particularly confined to the plane, reddening effects are expected to be less severe, and ii) since they constitute a minor fraction in the solar neighborhood, expanding the selection volume is necessary to secure a statistically robust sample for the CMD fitting. 

\end{enumerate}  

We adopted this nomenclature to explicitly distinguish our subsets from the previous work \citepalias[][]{emma2025} which incorporated a dependence with the distance to the Galactic plane, or from chemically defined disks. Both cylinders have been divided into the same 8 layers with the following boundaries:

\begin{equation}
           |Z|\; = [ 0.0 , 0.1 ,  0.2 , 0.3 , 0.4 , 0.5 , 0.6 , 0.8, 1.0]\; \text{kpc} \nonumber
\end{equation}

\section{Methodology}\label{section:methodology}
\subsection{Derivation of the Star Formation Histories}

We derive `dynamically evolved SFHs' (deSFHs, see below and \citetalias{Gallart2024} for a definition), using $\textit{CMDft.Gaia}$ for each layer in each subset separately. We employ the CMD fitting technique, that is, comparing observed CMDs with models obtained as a linear combination of simple stellar populations (SSPs) extracted from a `mother CMD' computed using stellar evolution models. The same parameters for the CMD fitting are used as in \citetalias{Gallart2024} and \citetalias{emma2025}. We adopted the BaSTI-IAC \citep{basti} stellar evolution library, assuming a binary fraction $\beta=0.3$, minimum mass ratio $q_\text{min}=0.1$, and a Kroupa initial mass function \citep[IMF,][]{kroupa1993}. Then, a simulation of the errors and completeness in each cylindrical layer is applied to the mother CMD, producing a mother-with-errors CMD for each layer. Finally, the algorithm searches for the combination of SSPs that best reproduces the observed CMD. An example of one fit along with the residuals is shown in Appendix~\ref{appendixCMDfit}.

$\textit{CMDft.Gaia}$ generates three outputs, each providing a complementary interpretation of the data:
    
\begin{enumerate}
    \item \textbf{deSFH}: This represents the total stellar mass per unit time and metallicity, that has been transformed into stars somewhere in the galaxy to account for the stars that are today in the studied volume. It explicitly accounts for the mass of stars that have already died, which are naturally missing from the observed CMD. 
    \item \textbf{solution CMD}: A synthetic stellar catalog that best reproduces the observed CMD, providing age and metallicity for each synthetic star. It is computed using the weights obtained by the deSFH on a mother CMD that includes simulated observational errors. For this reason, these CMDs are affected by the same photometric uncertainties and incompleteness as the actual \textit{Gaia} data. This catalogue is representative of the analyzed stars (i.e. observed by \textit{Gaia} and passing the quality criteria).
    \item \textbf{solution CMD$_\text{no-disp}$}: A synthetic stellar catalog generated using the same best-fit weights on an error-free, complete mother CMD. This catalog is immune to the observational biases introduced by data filtering. For instance, a reddening cut removes highly extincted stars from the observed CMD (and consequently in the same proportion from the mother CMD with errors). Since solution CMD$_\text{no-disp}$ bypasses these cuts, it provides a close representation of the true, intrinsic number of stars residing in the analyzed volume.
\end{enumerate}

\subsection{Correction to Obtain the Absolute Number of Stars} \label{section:corrections}

The fraction of stars removed by quality cuts is highly dependent on the vertical layer, being higher near the Galactic plane. To derive unbiased vertical profiles, we estimate the absolute number of stars in each age-metallicity bin by leveraging the aforementioned solution CMD$_\text{no-disp}$, and correcting for the fraction of stars that are removed from the observed and mother CMD by the absolute magnitude cuts applied. The fraction of stars fainter than the limiting magnitude varies across different SSPs. Therefore, as we aim to compare vertical profiles of different composite stellar populations, we must correct for this effect to avoid the bias that would be introduced due to the fixed magnitude cut. That is, we will take into account the faintest stars not considered in the fit and sometimes not even detected by \textit{Gaia}.  
    
For this correction, we calculated the fraction of stars fainter than the absolute magnitude cut for each SSP\footnote{We consider $M_\text{G} < 4.1$, as we have verified that the sample is complete at these magnitudes. Each SSP considered for the correction spans a bin size of $0.03$~Gyr and $0.007$~dex in age and metallicity, respectively, corresponding to a $400 \times 400$ grid.}, that is, we obtain the absolute number of stars ($N(\text{Age, [Z/H]})$). First, BaSTI-IAC isochrones are used to determine the mass limit corresponding to the magnitude cut for each SSP. Then, using a Kroupa IMF, the fraction is calculated as the ratio between the number of stars below our mass cut and the total population of currently living stars.

\subsection{Definition of qMAMs in the Age-Metallicity Plane}\label{methodology:qmams}

We aim to obtain vertical density profiles of the qMAMs, defined as populations of stars  that cluster together in the age-metallicity space. To define them, we apply the clustering algorithm \texttt{HDBSCAN}\footnote{Using the implementation of \texttt{Scikit-learn} \citep{scikit-learn}.} in the age-metallicity plane to the stars in the global solution CMD$_\text{no-disp}$ of each subset (i.e. the sum of solution CMDs$_\text{no-disp}$ of all layers). We obtain an age-metallicity map showing the stars classified into clusters. This map is then smoothed using a Gaussian filter to extract the contours that separate the different clusters identified by HDBSCAN, assigning all stars that fall within a specific boundary to the corresponding qMAM.

\begin{figure*}[!t]
   \centering
   \includegraphics[width=\textwidth]{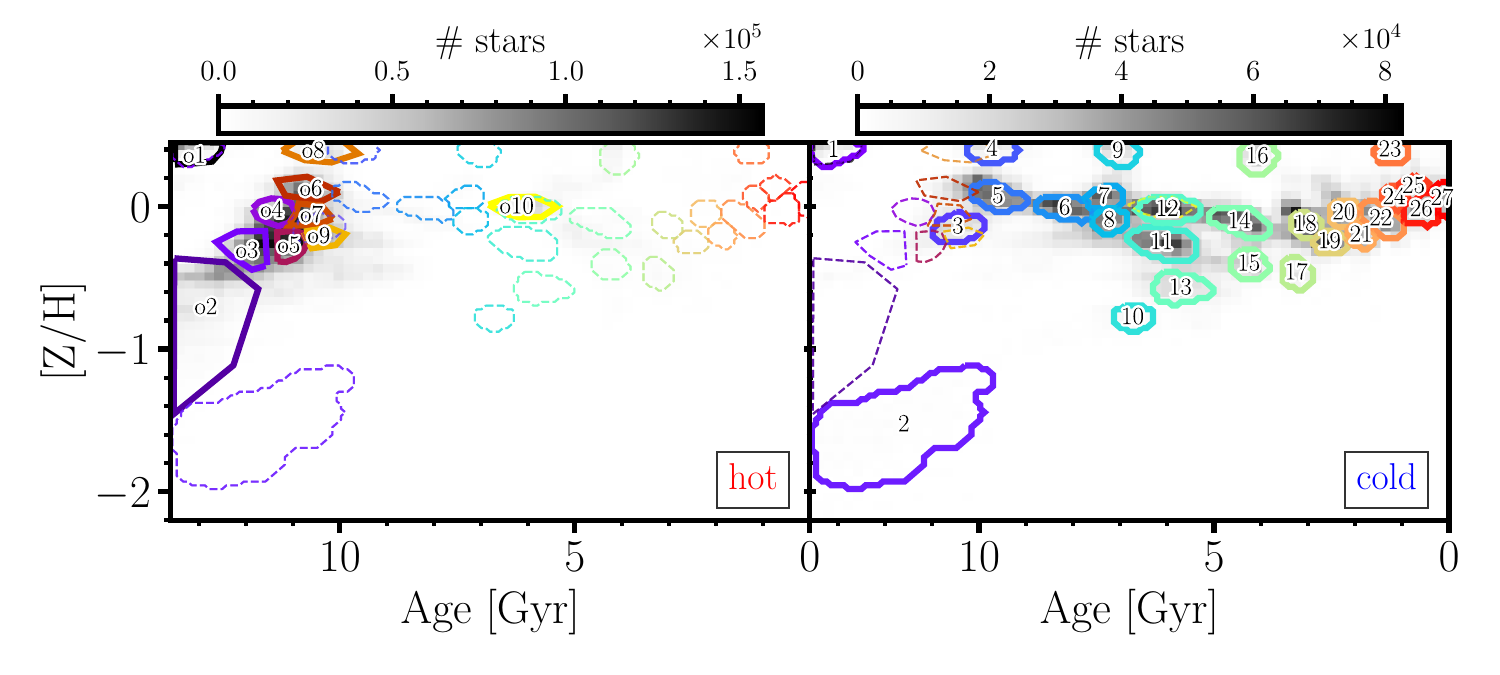}
      \caption{Definition of the qMAMs in the age-metallicity plane. The background colormap indicates the number of stars of the sum of all layers in each subset. Left panel: solid lines represent the density contours for qMAMs identified in the \texttt{hot} subset. Right panel: solid lines denote the qMAMs for the \texttt{cold} subset. To facilitate direct comparison, the contours from the opposite subset are overlaid in each panel as dashed lines.}
         \label{fig:contours}
\end{figure*}

Although we tried to define the qMAMs as automatically as possible to avoid bias, some human intervention was needed in the \texttt{hot} subset. The algorithm struggled to separate some populations: \texttt{HDBSCAN} identified the qMAMs \texttt{o3}, \texttt{o4}, \texttt{o5}, \texttt{o6}, \texttt{o7} and \texttt{o9} as a single one (see Figure~\ref{fig:contours}). By manually separating them, we verified that they have different vertical profiles, and thus we kept them separated to maximize the information (the conclusions drawn from this work remain unchanged if the age–metallicity plane is divided uniformly, see Appendix~\ref{appendixqMAMs}). Furthermore, in \citetalias{Gallart2024} we demonstrated that our age precision is $\sim 0.5$~Gyr at old ages, which allows us to make divisions of this size. For the \texttt{cold} subset, we retained the original clusters obtained by HDBSCAN.
    
Figure~\ref{fig:contours} shows the final contours for each qMAM, color-coded for clarity. The grayscale represents the number of stars in the age-metallicity plane in the global solution. The left and right panels display the distributions for the \texttt{hot} and \texttt{cold} subsets, respectively, with the qMAMs shown as solid lines. For a direct comparison, the contours of the opposite sample are overlaid as dashed lines. We obtained the absolute number of stars for each qMAM in each layer $i$ as the sum of the $N_i(\text{Age, [Z/H]})$ of bins included in the corresponding qMAM.

\begin{figure*}[!t]
   \centering
   \includegraphics[width=0.89\textwidth]{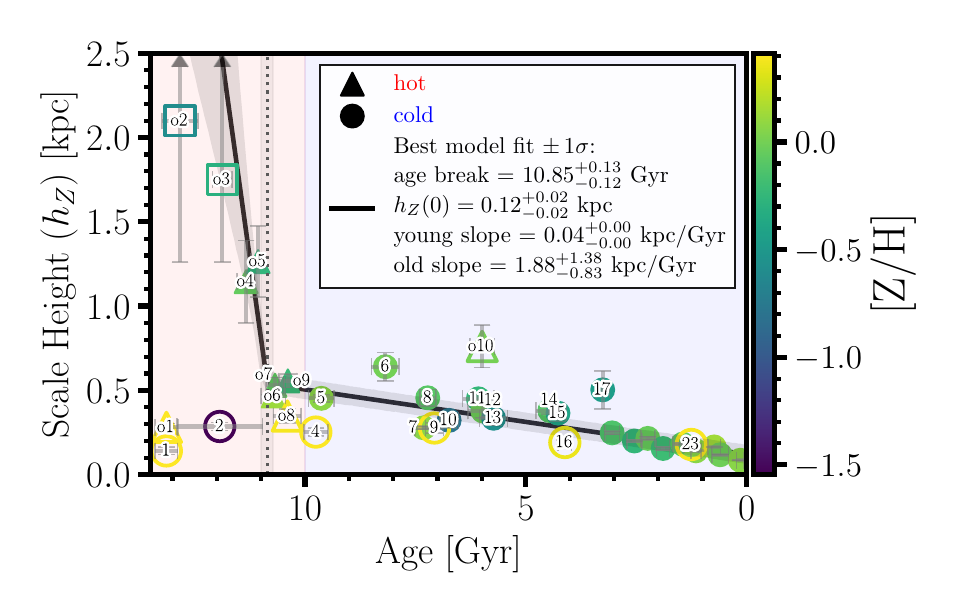}
      \caption{Vertical scale height, $h_Z$, vs. mean age for the qMAMs. Triangles (\texttt{hot}) and circles (\texttt{cold}) are color-coded by mean metallicity. For clarity, qMAM numbers (ordered sequentially by age) are omitted for younger populations. qMAM \texttt{3} is excluded due to relative $h_Z$ uncertainties $>30\%$. Although qMAMs \texttt{o2} and \texttt{o3} (square markers) also share large uncertainties and required special treatment (yielding rough $h_Z$ estimates, see Appendix~\ref{appendix:k2}), they are kept due to their relevance for the old population trend. The solid line and the grey shaded band represent the MCMC fit and its $1\sigma$ uncertainty, respectively, with the best-fitting $\pm1\sigma$ parameters given in the legend (see Appendix~\ref{appendix:mcmc} for details on the model). Open symbols denote the nine qMAMs excluded from the fit (see text).}
         \label{fig:scale_height}
\end{figure*}

\subsection{Vertical Profiles Fit}

For each qMAM in each layer $i$, we calculate the total stellar density ($\rho(\text{qMAM})$) by dividing the absolute number of stars in each qMAM by the total volume of each layer. This total density includes both the observed stars and those lost due to reddening/quality cuts and to the magnitude cut adopted for the SFH calculation.
    
Each of the vertical profiles has been fitted with a squared hyperbolic secant function, which represents the solution for an self-gravitating, isothermal gaseous disk, as originally derived by \citet{Spitzer}. It has the form:

\begin{equation}
    \rho_\text{fit}(|Z|, \text{qMAM}) = \rho_\text{fit}(0,\text{qMAM})\cdot\operatorname{sech}^2 \left(\dfrac{|Z|}{h_Z}\right)
\end{equation}
where $\rho_\text{fit}(|Z|,\text{qMAM})$ represents the fitted density profile, and $\rho_\text{fit}(0,\text{qMAM})$ and $h_Z$ are the central density and scale height, respectively. While both parameters are obtained from the fit, only $h_Z$ is considered in the analysis presented here.

\section{Results}\label{section:results}

Figure~\ref{fig:scale_height} illustrates the relationship between $h_Z$ and age\footnote{The mean age of each qMAM was calculated by weighting each grid bin in the age-metallicity plane by its absolute number of stars, $N(\text{qMAM})$, which we obtained by summing the counts over all layers.}. The $h_Z$ of qMAMs of the \texttt{hot} and the \texttt{cold} subsets are represented by triangles and circles, respectively. The vertical error bars denote the uncertainty derived from the $h_Z$ fitting procedure\footnote{These uncertainties correspond to the absolute 1$\sigma$ standard errors, which were estimated from the square root of the diagonal of the covariance matrix.}. The horizontal error bars represent the 1$\sigma$ standard deviation of stellar ages within each qMAM. Note that this spread does not correspond to the methodological uncertainty of the CMD fitting technique (see Figure~5 in \citetalias{Gallart2024}), but reflects the actual range of ages of each qMAM, combined with the CMD-fitting age precision. The color of each symbol represents its mean metallicity (calculated similarly to the mean age). 

The relation is modeled using a Markov chain Monte Carlo (MCMC) fit of a broken linear model composed of two linear trends, with the transition age allowed to vary. The best-fitting model is shown as a solid black line, with the $1\sigma$ uncertainty region shaded in grey. Details of the model and the MCMC analysis are provided in Appendix~\ref{appendix:mcmc}. A few qMAMs were excluded from the fit for physically motivated reasons. For the \texttt{hot} subset, we excluded \texttt{o10}, \texttt{o8}, \texttt{o1}, \texttt{o2} and \texttt{o3}. The first one was excluded because it is substantially younger than the rest of the \texttt{hot} populations, and it is completely disconnected from the rest of the populations in the age-metallicity plane.\footnote{In fact, this $\sim6$ Gyr old, solar metallicity population can be related to the \texttt{cold} disk qMAM \texttt{12}. The `hotter' stars of this population are systematically classified as \texttt{hot} disk, resulting in a higher scale height than their \texttt{cold} disk counterpart.} \texttt{o8} and \texttt{o1} are super metal-rich, and like qMAM \texttt{1} in the \texttt{cold} disk (of similar $h_Z$), they are populations likely migrated from the inner galaxy \citep{tomas2025}, and thus expected not to follow the same local $h_Z$ trend. qMAMs \texttt{1} and \texttt{o1} are, in fact, the same population split in both subsets. For the same reason, all super metal-rich qMAMs identified in the \texttt{cold} disk (\texttt{1}, \texttt{4}, \texttt{9}, \texttt{16} and \texttt{23}) are also excluded from the fit. In the case of \texttt{o2} and \texttt{o3}, the indicated $h_Z$ should be treated as a lower limit, as a precise value cannot be formally established (see Appendix~\ref{appendix:k2} for the special analysis of these particularly relevant qMAMs). For the \texttt{cold} subset, we included in the fit the qMAMs younger than 10 Gyr, as there is a clear separation between the \texttt{cold} and the \texttt{hot} sample at this age, see \citetalias{emma2025}. This excludes qMAM \texttt{1} and \texttt{2}, which have their own peculiarities and depart from the main trends\footnote{qMAM \texttt{2} is, surprisingly, only identified in the \texttt{cold} subset, with a very low $h_Z$. It is particularly intriguing that such an old and metal-poor population exhibits such a small scale height. The nature of this population is discussed in detail in A. B. Queiroz et al. (in preparation).}.
    
Figure~\ref{fig:scale_height} clearly shows that there is a correlation between $h_Z$ and age, with $h_Z$ decreasing for younger populations. However, beyond simply covering a different age range, the qMAMs in the \texttt{hot} subset follow a much steeper $h_Z$ trend compared with that of the \texttt{cold} one. 

~\\
~\\
~\\

It is interesting that qMAMs \texttt{o6}, \texttt{o7}, \texttt{o8} and \texttt{o9} remain consistent with the general trend observed in the \texttt{cold} subset. As these are the youngest qMAMs within the \texttt{hot} sample (except for \texttt{o10}), they can be interpreted as a transition population between \texttt{hot} and \texttt{cold} disks. However, the $h_Z$ of qMAMs \texttt{o2}, \texttt{o3}\footnote{While a formal fit could not be achieved, its nearly flat profile indicates a scale height exceeding those of the other \texttt{hot} disk populations. See Appendix~\ref{appendix:k2}.}, \texttt{o4} and \texttt{o5} are clearly higher than those of any other populations. 

In quantitative terms, the broken linear model yields a transition age of $10.85^{+0.13}_{-0.12}$ Gyr. The inferred settling rate is $1.88^{+1.38}_{-0.83}$ kpc/Gyr for the old-age regime and $0.038^{+0.004}_{-0.004}$ kpc/Gyr for the young part. Although the uncertainty associated with the old-age slope is relatively large, the results clearly indicate the presence of two distinct regimes with substantially different settling rates. Another noteworthy result is that the scale height predicted for newly formed stars is $0.12^{+0.02}_{-0.02}$ kpc.

It is precisely the fine age and metallicity resolution that allows us to capture these distinct regimes and resolve the earliest populations. This is key for revealing two distinct stages in the evolution of the disk thickness: a rapid thinning within the first $\simeq$ 2.5 Gyr, followed by a gradual one thereafter. This prominent change in slope indicates that the early disk was clearly thicker and hotter than at present.

\section{Discussion}\label{section:discussion}

\subsection{Phases of Milky Way assembly}

Several recent studies have capitalized on the extraordinary \textit{Gaia} datasets, complemented in some cases with data from ground-based spectroscopic surveys, to put on a firm observational ground the idea of a three-phase assembly history of the MW. This would include an initial turbulent pre-disk phase, primarily forming the low-metallicity halo both through in-situ star formation in the proto-Galaxy \citep[e.g., the Aurora population of][]{Belokurov2022} and accreted populations \citep{Gaia_Enceladus_HELMI2}, followed by two subsequent phases of {\it old} (thick) and {\it young} (thin) disk formation. The formation of the old disk is identified by a more \citep{Chandra2024} or less \citep{viswanathan} rapid transition from a velocity-dispersion dominated regime to one characterized by a substantial amount of net rotation, while a subsequent step to even larger orbital circularities marks the transition to the young disk. These works use metallicity as a proxy for age, even though they aim at disentangling {\it when} the different components emerged. The difficulty is twofold. First, several studies have shown that while a tight, well-defined age-metallicity relation exists for the thick disk (whether chemically, \citealt{Xiang22}, or kinematically, \citealt{Bensby2004}; \citetalias{emma2025}, defined), the corresponding relation for the thin disk is markedly flatter. Second, typical age uncertainties, even in the most exquisite age catalogues \citep[e.g.][]{Xiang22, Queiroz2023}, are still too large for a {\it fine temporal dissection} of the metallicity evolution of the thick disk. In spite of these relatively large individual age uncertainties, \citet{Feltzing2026} have recently placed a direct timescale on the emergence of a rotationally supported disk, dating the Galactic spin-up to $12.1 \pm 2.9$~Gyr (and $12.5 \pm 2.5$~Gyr for the high-$\alpha$ population) using subgiant stellar ages from \citet{Xiang22}.

The two-stage assembly of the MW {\it disk} is broadly reflected in the different global scale heights that are measured for the thin and thick disks~\citep{Juric2008, khana2025}. Additionally, a finer variation of the vertical distribution of different stellar populations as a function of their metallicity \citep{Bovy2012thick, Bovy2012spatial} and age \citep{Mackereth2017, Imig2025, Xiang2025} has been studied as a tracer of the mechanisms involved in disk formation and evolution. However, discerning between an \textit{upside-down} disk formation and \textit{secular dynamical heating} or other types of heating is proving to be particularly difficult. For example, recent observational studies, such as the one by \citet{Imig2025}, that dissects the disk populations taking into account both ages and abundances, finds that $h_Z$ smoothly increases with age and decreases with metallicity for the low-$\alpha$ populations, but while the $h_Z$ of the high-$\alpha$ population decreases with metallicity, a variation with age is not observed. They conclude that a combination of effects (\textit{dynamical heating} and older stars `born thicker') is responsible for the observed variations of $h_Z$ with age and metallicity, but without ascribing a dominance of these processes to a particular time-frame or evolutionary phase of the Galaxy.

Building upon a new observational framework, our results provide, for the first time, a measure of two distinct evolutionary phases, one earlier, and another one later than~$\simeq$~10.8~Gyr, clearly pointing to different physical mechanisms being dominant in each of these periods of MW evolution. As illustrated in Figure~\ref{fig:scale_height}, the qMAMs younger than 10.8 Gyr exhibit a progressive, gradual decrease in their $h_Z$ with time. This observed trend could be compatible with a \textit{secular dynamical heating} scenario, where stars are born in a dynamically cold gas layer and their orbits are subsequently and slowly heated over time. However, it is also consistent with an \textit{upside-down} formation scenario, in which the star-forming gas gradually cools and settles, and thus stars are born with progressively smaller vertical velocity dispersions. A more comprehensive view emerges when the behaviour of the older populations is incorporated into the picture. These qMAMs also exhibit a decrease of $h_Z$ with time, but at a significantly steeper rate compared to the younger counterpart. Viewing the system as a whole, a pure \textit{secular dynamical heating} scenario acting over the whole MW history becomes much less plausible. If dynamical heating were the sole mechanism at play, frequent mergers in the early Galaxy would naturally sustain a hot dynamical environment until the occurrence of the last major merger~$\sim$~10~Gyr~ago \citep[GSE,][]{Gaia_Enceladus_BELOKUROV, Gaia_Enceladus_HELMI, Gallart2019}. However, this is at odds with the extremely rapid settling observed in the oldest disk populations, with the typical $h_Z$ dropping from over 1.26~kpc to $\sim0.5$~kpc in just $\sim2$~Gyr and decreasing with increasing metallicity. This steep vertical settling of the early disk is fully compatible with the kinematic spin-up reported by \citet{Feltzing2026}, who suggested that the transition from a turbulent proto-Galaxy to a rotationally supported disk occurred extremely rapidly in the early MW. Our work strongly supports an \textit{upside-down} formation during the period these authors identify as the pre-spin-up and spin-up phase, in which the disk gradually settles close to the midplane.

Therefore this supports a view in which the evolution of the Galactic disk is intrinsically divided into distinct temporal regimes: an early epoch characterized by rapid assembly, followed by a subsequent, quiescent phase of evolution.
The oldest stars would be formed from a turbulent gas that cooled rapidly over time. Different physical mechanisms have been proposed to explain an early increased gas turbulence, such as frequent merger events \citep{Brook2004, Bird2013}, different modes of gas accretion \citep[see][]{gas2005, Belokurov2022}, or stellar feedback in a period of high star formation rate \citep{Navarro2018}.

\subsection{The theoretical framework}
This observed two-stage evolution of the Galactic disk is in agreement with that inferred in cosmological simulations \citep[e.g.][]{magicc}, thus supporting their adopted physical framework. In particular, \citet{Ma2017} found that in their simulated MW-mass galaxy, the vertical scale height exhibits a linearly increasing trend with age until $\sim 7$~Gyr, whereas the increase becomes drastically steeper for older populations (see their Figure~8). The authors interpret this sharp structural transition as a reflection of two fundamentally distinct modes of star formation and gas accretion. The oldest populations formed during a chaotic, highly turbulent phase dominated by rapid gas accretion, episodic starbursts, and mergers, resulting in stars born with large scale heights directly out of a gas-rich, dynamically hot medium. In contrast, as the galaxy evolved, the system stabilized, allowing for the formation of a rotationally supported gas disk. Stars formed during this subsequent secular phase initially exhibit much smaller scale heights. Furthermore, they demonstrate that the \textit{upside-down} formation and \textit{secular dynamical heating} scenarios act concurrently, with the latter contributing $30-40\%$ of the overall increase in scale height with stellar age. 

Finally, \citet{Gurvich23} analyzed MW-mass galaxies in the FIRE-2 suite \citep{FIRE2} and identified a `Pre-Disk Era' ($\gtrsim 8$~Gyr ago) when stars formed on turbulent, dispersion-dominated orbits. Crucially, they show that the large vertical extent and high velocity dispersion of these ancient populations were primarily imprinted at birth, rather than acquired through later secular heating. According to the authors, this early `bursty' star formation regime eventually evolves into a `steady' phase, leading to a rapid disk settling ($<1$ Gyr). During this period, the chaotic interstellar medium collapses into a coherently rotating thin disk. Similar conclusions were also reached by \citet{Bird2013} and \citet{Navarro2018}: while internal secular processes play a role, the structural vertical properties of stellar disks are fundamentally established early on by the conditions of the star-forming gas.

\subsection{On the thick and thin disk formation}
 
The MW disk is considered to be composed of two components, a thin and thick disk that have been defined either geometrically, kinematically or chemically, with the thick disk having a larger scale height, a higher velocity dispersion or being chemically enriched in $\alpha$-elements, respectively, compared to the thin disk. However, these different selections do not result in identical stellar samples, and this has led some authors to suggest that possibly the best definition would be by stellar age \citep{Bensby2014, kawata_chiappini}. In \citetalias{emma2025}, we selected thin and thick disk samples ($ks\_thick$, $ks\_thin$) with the classical criteria combining velocity dispersion and distances from the plane, it was shown that there was a separation of $ks\_thick$ and $ks\_thin$ stars selected under these criteria around 10 Gyr ago, thus indicating a kinematic transition at this time. In the same work, a comparison between the properties of the stars in the age-metallicity and the [$\alpha$/Fe]/[Z/H] planes, using [Z/H] to relate them, showed that the most metal-rich and younger stars in the $ks\_thick$ sample are already $\alpha$-poor, with the transition from $\alpha$-rich to $\alpha$-poor occurring in the metallicity interval $-0.35 < \rm{[Z/H]} < 0$ in which stars have ages between 12 and 10 Gyr old. This indicates that the transition from thick to thin disk chemical properties precedes the kinematic transition. Or, in other words, that the kinematic thick disk extends to ages younger than the chemical thick disk, at least in the range of galactocentric distances probed by the study.

In this Letter, the geometric and kinematic properties of stars in a similar volume as that studied in \citetalias{emma2025} are revealed in unprecedented detail, showing that $h_Z$ evolves with time at a drastically different rate before and after $\simeq$ 10.8 Gyr ago, with stars selected in the hot sample already showing $h_Z$ typical of the thin disk at this age. However, it is remarkable that populations just slightly older, $\simeq$ 11 Gyr, have $h_Z$ higher than that typically measured for the thick disk as a whole \citep[e.g. 0.9 kpc,][]{Juric2008}.\footnote{The fact that \citet{Juric2008} use an exponential fit does not affect the comparison. To verify it, we fitted our global samples using an exponential profile. This yields $h_Z$ of $0.88 \pm 0.04$~kpc and $0.27 \pm 0.01$~kpc for the \texttt{hot} and \texttt{cold} subsets, respectively, in excellent agreement with the canonical values of $0.9$ and $0.3$ kpc for the thick and thin disk.} 

The combination of all this information, which matches the predictions of leading cosmological simulations \citep{Ma2017, Gurvich23, Thulasidharan2026},  validates their broad adopted physical framework, providing clear evidence that suggest the following scenario for the formation of the MW disk. The chemical thick disk was formed by 11 Gyr ago \citep[in agreement with][ and references therein]{Queiroz2023}, during a period of intense star formation that managed to keep high both the $\alpha$-abundances and the gas turbulence, and to increase the overall metallicity to almost solar values. During this time-frame, the disk was gradually but rapidly becoming thinner. After this period of upside-down evolution, the MW disk transitioned to a much colder phase in which a steadier star formation rate kept the $\alpha$ abundances and overall metallicity close to the solar value.

\section{Conclusions}\label{section:conclusions}

We analyzed the vertical structure of the MW disk by obtaining the age-$h_Z$ relationship for simple populations (qMAMs), defined in age and metallicity and spanning a range of $Z$ near the solar radius. The age-metallicity distributions, of unprecedented precision across the entire lifetime of our Galaxy, have been obtained through CMD-fitting of Gaia data.

We find a clear correlation between stellar age and $h_Z$ across the disk in the solar neighborhood, with younger populations consistently exhibiting smaller $h_Z$. However, populations older than 10.8 Gyr follow a markedly different, faster $h_Z$ evolution compared to the younger ones. This indicates that the old (or thick) disk does not merely represent the old tail of a single continuous population, but rather a component with a distinct evolutionary history.

Enabled by precise ages and high completeness, our results exhibit two clear, time-dependent evolutionary phases for the Galactic disk which allow piecing together a comprehensive picture of the physical processes responsible for a two-phase assembly history of the Galactic disk. The first phase would be characterized by an \textit{upside-down} formation, in which highly turbulent star-forming gas gradually cools and settles over the course of just $\sim3$ Gyr, and thus stars are born with progressively smaller vertical velocity dispersion reflected in lower $h_Z$ despite this phase being still dominated by accretion. Once this chaotic phase concluded, incidentally and approximately after the GSE merger, the Galaxy transitioned to a more quiescent state, gradually cooling and settling to form the dynamically colder thin disk we observe today. The large $h_Z$ of the oldest populations are not merely the result of prolonged \textit{secular dynamical heating}, but a structural fossil of the early Galaxy's assembly phase.
    
%%%%%%%%%%%%%%%%%%%%%%%%%%%%%%%%%%%%%%%%%%%%%%%%%%%%%%%%%%%%%%

\begin{acknowledgements}
DM and CG acknowledge support from the Agencia Estatal de Investigación del Ministerio de Ciencia e Innovación (AEI-MCINN) under grant `At the forefront of Galactic Archaeology: evolution of the luminous and dark matter components of the MW and LG dwarf galaxies in the {\it Gaia} era, with reference PID2023-150319NB-C21/10.13039/501100011033. EFA acknowledges support from the Agencia Estatal de Investigación del Ministerio de Ciencia e Innovación (AEI-MCINN) under grant `At the forefront of Galactic Archaeology: the chemical aspects' with reference PID2023-150319NB-C22/10.13039/501100011033. DM, CG, GAM and PAP acknowledge financial support from the European Research Council (ERC-2024-AdG `ChronoGal: Chronology of our Galaxy from Gaia CMD-fitting', Grant Agreement 101201412). Views and opinions expressed are, however, those of the author(s) only and do not necessarily reflect those of the European Union or the  European Research Council Executive Agency (ERCEA). Neither the European Union nor the granting authority can be held responsible for them. DM acknowledges support from the European Union (Widening Participation, ExGal-Twin, GA 101158446). TRL acknowledges support from Ramón y Cajal fellowship (RYC2023-043063-I, financed by MCIU/AEI/10.13039/501100011033 and by the FSE+) as well as research projects PID2023-149578NB-I00 and PID2023-150178NB-I00. ABAQ acknowledges support from the Juan de la Cierva programme (grant JDC2023-052806-I), funded by MICIU/AEI/10.13039/501100011033 and by the European Social Fund Plus (FSE+). SC acknowledges support from the project ``ASTRA: Across Space and Time: Relics \& Archaeology'' (P.I. M. Marconi) funded by INAF-Instituto Nazionale di Astrofisica. YGK acknowledges financial support from PREP2023-001684 funded by MCIU/AEI/10.13039/501100011033 and the FSE+. JFN acknowledges support from the Fundación Occident and the Instituto de Astrofísica de Canarias under the Visiting Researcher Programme 2025-2028 agreed between both institutions. The authors wish to acknowledge the contribution of the IAC High-Performance Computing support team and hardware facilities to the results of this research. This work is part of grant CEX2025-001609-S, awarded to the Instituto de Astrofísica de Canarias under the Severo Ochoa Centre of Excellence program and funded by MICIU/AEI/10.13039/501100011033. This work used data from the European Space Agency (ESA) mission {\it Gaia} (\url{https://www.cosmos.esa.int/gaia}), processed by the {\it Gaia} Data Processing and Analysis Consortium (DPAC, \url{https://www.cosmos.esa.int/web/gaia/dpac/consortium}). Funding for the DPAC has been provided by national institutions, in particular the institutions participating in the {\it Gaia} Multilateral Agreement.
\end{acknowledgements}

%%%%%%%%%%%%%%%%%%%%%%%%%%%%%%%%%%%%%%%%%%%%%%%%%%%%%%%%%%%%%%

\bibliographystyle{aasjournalv7}
\bibliography{bibliography}

@article{Chandra2024,
   author = {Vedant Chandra and Vadim A. Semenov and Hans-Walter Rix and Charlie Conroy and Ana Bonaca and Rohan P. Naidu and René Andrae and Jiadong 佳东 Li 李 and Lars Hernquist},
   doi = {10.3847/1538-4357/ad5b60},
   issn = {0004-637X},
   issue = {1},
   journal = {The Astrophysical Journal},
   month = {9},
   pages = {112},
   title = {The Three-phase Evolution of the Milky Way},
   volume = {972},
   year = {2024},
adsurl = {https://ui.adsabs.harvard.edu/abs/2024ApJ...972..112C}
}

@article{Gallart2024,
   author = {Carme Gallart and Francisco Surot and Santi Cassisi and Emma Fernández-Alvar and David Mirabal and Alicia Rivero and Tomás Ruiz-Lara and Judith Santos-Torres and Guillem Aznar-Menargues and Giuseppina Battaglia and Anna B. Queiroz and Matteo Monelli and Eugene Vasiliev and Cristina Chiappini and Amina Helmi and Vanessa Hill and Davide Massari and Guillaume F. Thomas},
   doi = {10.1051/0004-6361/202349078},
   issn = {0004-6361},
   journal = {\aap},
   month = {7},
   pages = {A168},
   title = {Chronology of our Galaxy from <i>Gaia</i> colour–magnitude diagram fitting (ChronoGal)},
   volume = {687},
   year = {2024},
adsurl = {https://ui.adsabs.harvard.edu/abs/2024A&A...687A.168G}
    
}

@article{Gallart2019,
   author = {Carme Gallart and Edouard J. Bernard and Chris B. Brook and Tomás Ruiz-Lara and Santi Cassisi and Vanessa Hill and Matteo Monelli},
   doi = {10.1038/s41550-019-0829-5},
   issn = {2397-3366},
   issue = {10},
   journal = {Nature Astronomy},
   month = {7},
   pages = {932-939},
   title = {Uncovering the birth of the Milky Way through accurate stellar ages with Gaia},
   volume = {3},
   year = {2019},
adsurl = {https://ui.adsabs.harvard.edu/abs/2019NatAs...3..932G}
}

@ARTICLE{magicc,
       author = {{Stinson}, G.~S. and {Brook}, C. and {Macci{\`o}}, A.~V. and {Wadsley}, J. and {Quinn}, T.~R. and {Couchman}, H.~M.~P.},
        title = "{Making Galaxies In a Cosmological Context: the need for early stellar feedback}",
      journal = {\mnras},
         year = 2013,
        month = jan,
       volume = {428},
       number = {1},
        pages = {129-140},
          doi = {10.1093/mnras/sts028},
archivePrefix = {arXiv},
       eprint = {1208.0002},
 primaryClass = {astro-ph.CO},
       adsurl = {https://ui.adsabs.harvard.edu/abs/2013MNRAS.428..129S}
}

@ARTICLE{Gaia_Enceladus_BELOKUROV,
       author = {{Belokurov}, V. and {Erkal}, D. and {Evans}, N.~W. and {Koposov}, S.~E. and {Deason}, A.~J.},
        title = "{Co-formation of the disc and the stellar halo}",
      journal = {\mnras},
         year = 2018,
        month = jul,
       volume = {478},
       number = {1},
        pages = {611-619},
          doi = {10.1093/mnras/sty982},
archivePrefix = {arXiv},
       eprint = {1802.03414},
 primaryClass = {astro-ph.GA},
       adsurl = {https://ui.adsabs.harvard.edu/abs/2018MNRAS.478..611B}
}

@ARTICLE{Gaia_Enceladus_HELMI,
       author = {{Helmi}, Amina and {Babusiaux}, Carine and {Koppelman}, Helmer H. and {Massari}, Davide and {Veljanoski}, Jovan and {Brown}, Anthony G.~A.},
        title = "{The merger that led to the formation of the Milky Way's inner stellar halo and thick disk}",
      journal = {\nat},
         year = 2018,
        month = oct,
       volume = {563},
       number = {7729},
        pages = {85-88},
          doi = {10.1038/s41586-018-0625-x},
archivePrefix = {arXiv},
       eprint = {1806.06038},
 primaryClass = {astro-ph.GA},
       adsurl = {https://ui.adsabs.harvard.edu/abs/2018Natur.563...85H}
}

@ARTICLE{Gaia_Enceladus_HELMI2,
       author = {{Helmi}, Amina},
        title = "{Streams, Substructures, and the Early History of the Milky Way}",
      journal = {\araa},
         year = 2020,
        month = aug,
       volume = {58},
        pages = {205-256},
          doi = {10.1146/annurev-astro-032620-021917},
archivePrefix = {arXiv},
       eprint = {2002.04340},
 primaryClass = {astro-ph.GA},
       adsurl = {https://ui.adsabs.harvard.edu/abs/2020ARA&A..58..205H}
}

@ARTICLE{gaiadr3,
       author = {{Gaia Collaboration} and {Vallenari}, A. and {Brown}, A.~G.~A. and {Prusti}, T. and {de Bruijne}, J.~H.~J. and {Arenou}, F. and {Babusiaux}, C. and {Biermann}, M. and {Creevey}, O.~L. and {Ducourant}, C. and {Evans}, D.~W. and {Eyer}, L. and {Guerra}, R. and {Hutton}, A. and {Jordi}, C. and {Klioner}, S.~A. and {Lammers}, U.~L. and {Lindegren}, L. and {Luri}, X. and {Mignard}, F. and {Panem}, C. and {Pourbaix}, D. and {Randich}, S. and {Sartoretti}, P. and {Soubiran}, C. and {Tanga}, P. and {Walton}, N.~A. and {Bailer-Jones}, C.~A.~L. and {Bastian}, U. and {Drimmel}, R. and {Jansen}, F. and {Katz}, D. and {Lattanzi}, M.~G. and {van Leeuwen}, F. and {Bakker}, J. and {Cacciari}, C. and {Casta{\~n}eda}, J. and {De Angeli}, F. and {Fabricius}, C. and {Fouesneau}, M. and {Fr{\'e}mat}, Y. and {Galluccio}, L. and {Guerrier}, A. and {Heiter}, U. and {Masana}, E. and {Messineo}, R. and {Mowlavi}, N. and {Nicolas}, C. and {Nienartowicz}, K. and {Pailler}, F. and {Panuzzo}, P. and {Riclet}, F. and {Roux}, W. and {Seabroke}, G.~M. and {Sordo}, R. and {Th{\'e}venin}, F. and {Gracia-Abril}, G. and {Portell}, J. and {Teyssier}, D. and {Altmann}, M. and {Andrae}, R. and {Audard}, M. and {Bellas-Velidis}, I. and {Benson}, K. and {Berthier}, J. and {Blomme}, R. and {Burgess}, P.~W. and {Busonero}, D. and {Busso}, G. and {C{\'a}novas}, H. and {Carry}, B. and {Cellino}, A. and {Cheek}, N. and {Clementini}, G. and {Damerdji}, Y. and {Davidson}, M. and {de Teodoro}, P. and {Nu{\~n}ez Campos}, M. and {Delchambre}, L. and {Dell'Oro}, A. and {Esquej}, P. and {Fern{\'a}ndez-Hern{\'a}ndez}, J. and {Fraile}, E. and {Garabato}, D. and {Garc{\'\i}a-Lario}, P. and {Gosset}, E. and {Haigron}, R. and {Halbwachs}, J. -L. and {Hambly}, N.~C. and {Harrison}, D.~L. and {Hern{\'a}ndez}, J. and {Hestroffer}, D. and {Hodgkin}, S.~T. and {Holl}, B. and {Jan{\ss}en}, K. and {Jevardat de Fombelle}, G. and {Jordan}, S. and {Krone-Martins}, A. and {Lanzafame}, A.~C. and {L{\"o}ffler}, W. and {Marchal}, O. and {Marrese}, P.~M. and {Moitinho}, A. and {Muinonen}, K. and {Osborne}, P. and {Pancino}, E. and {Pauwels}, T. and {Recio-Blanco}, A. and {Reyl{\'e}}, C. and {Riello}, M. and {Rimoldini}, L. and {Roegiers}, T. and {Rybizki}, J. and {Sarro}, L.~M. and {Siopis}, C. and {Smith}, M. and {Sozzetti}, A. and {Utrilla}, E. and {van Leeuwen}, M. and {Abbas}, U. and {{\'A}brah{\'a}m}, P. and {Abreu Aramburu}, A. and {Aerts}, C. and {Aguado}, J.~J. and {Ajaj}, M. and {Aldea-Montero}, F. and {Altavilla}, G. and {{\'A}lvarez}, M.~A. and {Alves}, J. and {Anders}, F. and {Anderson}, R.~I. and {Anglada Varela}, E. and {Antoja}, T. and {Baines}, D. and {Baker}, S.~G. and {Balaguer-N{\'u}{\~n}ez}, L. and {Balbinot}, E. and {Balog}, Z. and {Barache}, C. and {Barbato}, D. and {Barros}, M. and {Barstow}, M.~A. and {Bartolom{\'e}}, S. and {Bassilana}, J. -L. and {Bauchet}, N. and {Becciani}, U. and {Bellazzini}, M. and {Berihuete}, A. and {Bernet}, M. and {Bertone}, S. and {Bianchi}, L. and {Binnenfeld}, A. and {Blanco-Cuaresma}, S. and {Blazere}, A. and {Boch}, T. and {Bombrun}, A. and {Bossini}, D. and {Bouquillon}, S. and {Bragaglia}, A. and {Bramante}, L. and {Breedt}, E. and {Bressan}, A. and {Brouillet}, N. and {Brugaletta}, E. and {Bucciarelli}, B. and {Burlacu}, A. and {Butkevich}, A.~G. and {Buzzi}, R. and {Caffau}, E. and {Cancelliere}, R. and {Cantat-Gaudin}, T. and {Carballo}, R. and {Carlucci}, T. and {Carnerero}, M.~I. and {Carrasco}, J.~M. and {Casamiquela}, L. and {Castellani}, M. and {Castro-Ginard}, A. and {Chaoul}, L. and {Charlot}, P. and {Chemin}, L. and {Chiaramida}, V. and {Chiavassa}, A. and {Chornay}, N. and {Comoretto}, G. and {Contursi}, G. and {Cooper}, W.~J. and {Cornez}, T. and {Cowell}, S. and {Crifo}, F. and {Cropper}, M. and {Crosta}, M. and {Crowley}, C. and {Dafonte}, C. and {Dapergolas}, A. and {David}, M. and {David}, P. and {de Laverny}, P. and {De Luise}, F. and {De March}, R.},
        title = "{Gaia Data Release 3. Summary of the content and survey properties}",
      journal = {\aap},
         year = 2023,
        month = jun,
       volume = {674},
          eid = {A1},
        pages = {A1},
          doi = {10.1051/0004-6361/202243940},
archivePrefix = {arXiv},
       eprint = {2208.00211},
 primaryClass = {astro-ph.GA},
       adsurl = {https://ui.adsabs.harvard.edu/abs/2023A&A...674A...1G}
}

@ARTICLE{emma2025,
       author = {{Fern{\'a}ndez-Alvar}, Emma and {Ruiz-Lara}, Tom{\'a}s and {Gallart}, Carme and {Cassisi}, Santi and {Surot}, Francisco and {Gonz{\'a}lez-Koda}, Yllari K. and {Callingham}, Thomas M. and {Queiroz}, Anna B. and {Battaglia}, Giuseppina and {Thomas}, Guillaume and {Chiappini}, Cristina and {Hill}, Vanessa and {Dodd}, Emma and {Helmi}, Amina and {Aznar-Menargues}, Guillem and {de la Cueva}, Alejandro and {Mirabal}, David and {Quintana-Ansaldo}, M{\'o}nica and {Rivero}, Alicia},
        title = "{Chronology of our Galaxy from Gaia colour─magnitude diagram fitting (ChronoGal): II. Unveiling the formation and evolution of the kinematically selected thick and thin discs}",
      journal = {\aap},
         year = 2025,
        month = dec,
       volume = {704},
          eid = {A258},
        pages = {A258},
          doi = {10.1051/0004-6361/202553814},
archivePrefix = {arXiv},
       eprint = {2503.19536},
 primaryClass = {astro-ph.GA},
       adsurl = {https://ui.adsabs.harvard.edu/abs/2025A&A...704A.258F}
}

@ARTICLE{l22,
       author = {{Lallement}, R. and {Vergely}, J.~L. and {Babusiaux}, C. and {Cox}, N.~L.~J.},
        title = "{Updated Gaia-2MASS 3D maps of Galactic interstellar dust}",
      journal = {\aap},
         year = 2022,
        month = may,
       volume = {661},
          eid = {A147},
        pages = {A147},
          doi = {10.1051/0004-6361/202142846},
archivePrefix = {arXiv},
       eprint = {2203.01627},
 primaryClass = {astro-ph.GA},
       adsurl = {https://ui.adsabs.harvard.edu/abs/2022A&A...661A.147L}
}

@ARTICLE{kawata_chiappini,
       author = {{Kawata}, D. and {Chiappini}, C.},
        title = "{Milky Way's thick and thin disk: Is there a distinct thick disk?}",
      journal = {Astronomische Nachrichten},
         year = 2016,
        month = sep,
       volume = {337},
       number = {8-9},
        pages = {976},
          doi = {10.1002/asna.201612421},
archivePrefix = {arXiv},
       eprint = {1608.01698},
 primaryClass = {astro-ph.GA},
       adsurl = {https://ui.adsabs.harvard.edu/abs/2016AN....337..976K}
}

@ARTICLE{viswanathan,
       author = {{Viswanathan}, Akshara and {Horta}, Danny and {Price-Whelan}, Adrian M. and {Starkenburg}, Else},
        title = "{A slow spin to win: The gradual kinematic evolution across metallicities of the proto-Galaxy to the high-{\ensuremath{\alpha}} disc}",
      journal = {\aap},
         year = 2025,
        month = nov,
       volume = {703},
          eid = {A183},
        pages = {A183},
          doi = {10.1051/0004-6361/202453063},
archivePrefix = {arXiv},
       eprint = {2411.12165},
 primaryClass = {astro-ph.GA},
       adsurl = {https://ui.adsabs.harvard.edu/abs/2025A&A...703A.183V}
}

@ARTICLE{Ramirez2013,
       author = {{Ram{\'\i}rez}, I. and {Allende Prieto}, C. and {Lambert}, D.~L.},
        title = "{Oxygen Abundances in Nearby FGK Stars and the Galactic Chemical Evolution of the Local Disk and Halo}",
      journal = {\apj},
         year = 2013,
        month = feb,
       volume = {764},
       number = {1},
          eid = {78},
        pages = {78},
          doi = {10.1088/0004-637X/764/1/78},
archivePrefix = {arXiv},
       eprint = {1301.1582},
 primaryClass = {astro-ph.SR},
       adsurl = {https://ui.adsabs.harvard.edu/abs/2013ApJ...764...78R}
}

@ARTICLE{Reid2019,
       author = {{Reid}, M.~J. and {Menten}, K.~M. and {Brunthaler}, A. and {Zheng}, X.~W. and {Dame}, T.~M. and {Xu}, Y. and {Li}, J. and {Sakai}, N. and {Wu}, Y. and {Immer}, K. and {Zhang}, B. and {Sanna}, A. and {Moscadelli}, L. and {Rygl}, K.~L.~J. and {Bartkiewicz}, A. and {Hu}, B. and {Quiroga-Nu{\~n}ez}, L.~H. and {van Langevelde}, H.~J.},
        title = "{Trigonometric Parallaxes of High-mass Star-forming Regions: Our View of the Milky Way}",
      journal = {\apj},
         year = 2019,
        month = nov,
       volume = {885},
       number = {2},
          eid = {131},
        pages = {131},
          doi = {10.3847/1538-4357/ab4a11},
archivePrefix = {arXiv},
       eprint = {1910.03357},
 primaryClass = {astro-ph.GA},
       adsurl = {https://ui.adsabs.harvard.edu/abs/2019ApJ...885..131R}
}

@ARTICLE{Soubiran2003,
       author = {{Soubiran}, C. and {Bienaym{\'e}}, O. and {Siebert}, A.},
        title = "{Vertical distribution of Galactic disk stars.  I. Kinematics and metallicity}",
      journal = {\aap},
         year = 2003,
        month = jan,
       volume = {398},
        pages = {141-151},
          doi = {10.1051/0004-6361:20021615},
archivePrefix = {arXiv},
       eprint = {astro-ph/0210628},
 primaryClass = {astro-ph},
       adsurl = {https://ui.adsabs.harvard.edu/abs/2003A&A...398..141S}
}

@ARTICLE{Chiba2000,
       author = {{Chiba}, Masashi and {Beers}, Timothy C.},
        title = "{Kinematics of Metal-poor Stars in the Galaxy. III. Formation of the Stellar Halo and Thick Disk as Revealed from a Large Sample of Nonkinematically Selected Stars}",
      journal = {\aj},
         year = 2000,
        month = jun,
       volume = {119},
       number = {6},
        pages = {2843-2865},
          doi = {10.1086/301409},
archivePrefix = {arXiv},
       eprint = {astro-ph/0003087},
 primaryClass = {astro-ph},
       adsurl = {https://ui.adsabs.harvard.edu/abs/2000AJ....119.2843C}
}

@ARTICLE{kroupa1993,
       author = {{Kroupa}, Pavel and {Tout}, Christopher A. and {Gilmore}, Gerard},
        title = "{The Distribution of Low-Mass Stars in the Galactic Disc}",
      journal = {\mnras},
         year = 1993,
        month = jun,
       volume = {262},
        pages = {545-587},
          doi = {10.1093/mnras/262.3.545},
       adsurl = {https://ui.adsabs.harvard.edu/abs/1993MNRAS.262..545K}
}

@ARTICLE{yllari2025,
       author = {{Gonz{\'a}lez-Koda}, Yllari K. and {Ruiz-Lara}, Tom{\'a}s and {Gallart}, Carme and {Ceccarelli}, Edoardo and {Dodd}, Emma and {Fern{\'a}ndez-Alvar}, Emma and {Cassisi}, Santi and {Surot}, Francisco and {Aguado-Agelet}, Fernando and {Massari}, Davide and {Monelli}, Matteo and {Callingham}, Thomas M. and {Helmi}, Amina and {Aznar-Menargues}, Guillem and {Mirabal}, David and {Rivero}, Alicia and {Queiroz}, Anna B.},
        title = "{Chronology of our Galaxy from Gaia colour-magnitude diagram fitting (ChronoGal): III. Age and metallicity distribution of Gaia-Sausage-Enceladus stars near the Sun}",
      journal = {\aap},
         year = 2025,
        month = dec,
       volume = {704},
          eid = {A259},
        pages = {A259},
          doi = {10.1051/0004-6361/202554272},
archivePrefix = {arXiv},
       eprint = {2502.20439},
 primaryClass = {astro-ph.GA},
       adsurl = {https://ui.adsabs.harvard.edu/abs/2025A&A...704A.259G}
}

@ARTICLE{tomas2025,
       author = {{Ruiz-Lara}, Tom{\'a}s and {Mirabal}, David and {Gallart}, Carme and {Grand}, Robert and {Fragkoudi}, Francesca and {P{\'e}rez}, Isabel and {Cassisi}, Santi and {Fern{\'a}ndez-Alvar}, Emma and {Queiroz}, Anna B. and {Aznar-Menargues}, Guillem and {Gonz{\'a}lez-Koda}, Yllari K. and {Rivero}, Alicia and {Surot}, Francisco and {Thomas}, Guillaume F. and {Bieri}, Rebekka and {Gomez}, Facundo A. and {Pakmor}, R{\"u}diger and {van de Voort}, Freeke},
        title = "{Chronology of our Galaxy from Gaia colour-magnitude diagram fitting (ChronoGal): IV. The inner Milky Way stellar age distribution}",
      journal = {\aap},
         year = 2026,
        month = jan,
       volume = {705},
          eid = {A92},
        pages = {A92},
          doi = {10.1051/0004-6361/202556468},
archivePrefix = {arXiv},
       eprint = {2510.02238},
 primaryClass = {astro-ph.GA},
       adsurl = {https://ui.adsabs.harvard.edu/abs/2026A&A...705A..92R}
}

@ARTICLE{basti,
       author = {{Hidalgo}, Sebastian L. and {Pietrinferni}, Adriano and {Cassisi}, Santi and {Salaris}, Maurizio and {Mucciarelli}, Alessio and {Savino}, Alessandro and {Aparicio}, Antonio and {Silva Aguirre}, Victor and {Verma}, Kuldeep},
        title = "{The Updated BaSTI Stellar Evolution Models and Isochrones. I. Solar-scaled Calculations}",
      journal = {\apj},
         year = 2018,
        month = apr,
       volume = {856},
       number = {2},
          eid = {125},
        pages = {125},
          doi = {10.3847/1538-4357/aab158},
archivePrefix = {arXiv},
       eprint = {1802.07319},
 primaryClass = {astro-ph.GA},
       adsurl = {https://ui.adsabs.harvard.edu/abs/2018ApJ...856..125H}
}

@ARTICLE{scikit-learn,
       author = {{Pedregosa}, Fabian and {Varoquaux}, Ga{\"e}l and {Gramfort}, Alexandre and {Michel}, Vincent and {Thirion}, Bertrand and {Grisel}, Olivier and {Blondel}, Mathieu and {M{\"u}ller}, Andreas and {Nothman}, Joel and {Louppe}, Gilles and {Prettenhofer}, Peter and {Weiss}, Ron and {Dubourg}, Vincent and {Vanderplas}, Jake and {Passos}, Alexandre and {Cournapeau}, David and {Brucher}, Matthieu and {Perrot}, Matthieu and {Duchesnay}, {\'E}douard},
        title = "{Scikit-learn: Machine Learning in Python}",
      journal = {Journal of Machine Learning Research},
         year = 2011,
        month = oct,
       volume = {12},
        pages = {2825-2830},
          doi = {10.48550/arXiv.1201.0490},
archivePrefix = {arXiv},
       eprint = {1201.0490},
 primaryClass = {cs.LG},
       adsurl = {https://ui.adsabs.harvard.edu/abs/2011JMLR...12.2825P}
}

@ARTICLE{Spitzer,
       author = {{Spitzer}, Jr., Lyman},
        title = "{The Dynamics of the Interstellar Medium. III. Galactic Distribution.}",
      journal = {\apj},
         year = 1942,
        month = may,
       volume = {95},
        pages = {329},
          doi = {10.1086/144407},
       adsurl = {https://ui.adsabs.harvard.edu/abs/1942ApJ....95..329S}
}

@ARTICLE{Imig2025,
       author = {{Imig}, Julie and {Holtzman}, Jon A. and {Zasowski}, Gail and {Lian}, Jianhui and {Boardman}, Nicholas F. and {Stone-Martinez}, Alexander and {Mackereth}, J. Ted and {Prescott}, Moire K.~M. and {Beaton}, Rachael L. and {Beers}, Timothy C. and {Bizyaev}, Dmitry and {Blanton}, Michael R. and {Cunha}, Katia and {Fern{\'a}ndez-Trincado}, Jos{\'e} G. and {Fielder}, Catherine E. and {Hasselquist}, Sten and {Hayes}, Christian R. and {Haywood}, Misha and {J{\"o}nsson}, Henrik and {Lane}, Richard R. and {Majewski}, Steven R. and {M{\'e}sz{\'a}ros}, Szabolcs and {Minchev}, Ivan and {Nidever}, David L. and {Nitschelm}, Christian and {Sobeck}, Jennifer},
        title = "{A Galactic Self-portrait: Density Structure and Integrated Properties of the Milky Way Disk}",
      journal = {\apj},
         year = 2025,
        month = sep,
       volume = {990},
       number = {2},
          eid = {203},
        pages = {203},
          doi = {10.3847/1538-4357/adf723},
archivePrefix = {arXiv},
       eprint = {2507.17629},
 primaryClass = {astro-ph.GA},
       adsurl = {https://ui.adsabs.harvard.edu/abs/2025ApJ...990..203I}
}

@ARTICLE{Gilmore1983,
       author = {{Gilmore}, G. and {Reid}, N.},
        title = "{New light on faint stars - III. Galactic structure towards the South Pole and the Galactic thick disc.}",
      journal = {\mnras},
         year = 1983,
        month = mar,
       volume = {202},
        pages = {1025-1047},
          doi = {10.1093/mnras/202.4.1025},
       adsurl = {https://ui.adsabs.harvard.edu/abs/1983MNRAS.202.1025G}
}

@ARTICLE{Bovy2012spatial,
       author = {{Bovy}, Jo and {Rix}, Hans-Walter and {Liu}, Chao and {Hogg}, David W. and {Beers}, Timothy C. and {Lee}, Young Sun},
        title = "{The Spatial Structure of Mono-abundance Sub-populations of the Milky Way Disk}",
      journal = {\apj},
         year = 2012,
        month = jul,
       volume = {753},
       number = {2},
          eid = {148},
        pages = {148},
          doi = {10.1088/0004-637X/753/2/148},
archivePrefix = {arXiv},
       eprint = {1111.1724},
 primaryClass = {astro-ph.GA},
       adsurl = {https://ui.adsabs.harvard.edu/abs/2012ApJ...753..148B}
}

@ARTICLE{Bovy2012thick,
       author = {{Bovy}, Jo and {Rix}, Hans-Walter and {Hogg}, David W.},
        title = "{The Milky Way Has No Distinct Thick Disk}",
      journal = {\apj},
         year = 2012,
        month = jun,
       volume = {751},
       number = {2},
          eid = {131},
        pages = {131},
          doi = {10.1088/0004-637X/751/2/131},
archivePrefix = {arXiv},
       eprint = {1111.6585},
 primaryClass = {astro-ph.GA},
       adsurl = {https://ui.adsabs.harvard.edu/abs/2012ApJ...751..131B}
}

@ARTICLE{Mackereth2017,
       author = {{Mackereth}, J. Ted and {Bovy}, Jo and {Schiavon}, Ricardo P. and {Zasowski}, Gail and {Cunha}, Katia and {Frinchaboy}, Peter M. and {Garc{\'\i}a Perez}, Ana E. and {Hayden}, Michael R. and {Holtzman}, Jon and {Majewski}, Steven R. and {M{\'e}sz{\'a}ros}, Szabolcs and {Nidever}, David L. and {Pinsonneault}, Marc and {Shetrone}, Matthew D.},
        title = "{The age-metallicity structure of the Milky Way disc using APOGEE}",
      journal = {\mnras},
         year = 2017,
        month = nov,
       volume = {471},
       number = {3},
        pages = {3057-3078},
          doi = {10.1093/mnras/stx1774},
archivePrefix = {arXiv},
       eprint = {1706.00018},
 primaryClass = {astro-ph.GA},
       adsurl = {https://ui.adsabs.harvard.edu/abs/2017MNRAS.471.3057M}
}

@ARTICLE{Brook2012,
       author = {{Brook}, C.~B. and {Stinson}, G. and {Gibson}, B.~K. and {Wadsley}, J. and {Quinn}, T.},
        title = "{MaGICC discs: matching observed galaxy relationships over a wide stellar mass range}",
      journal = {\mnras},
         year = 2012,
        month = aug,
       volume = {424},
       number = {2},
        pages = {1275-1283},
          doi = {10.1111/j.1365-2966.2012.21306.x},
archivePrefix = {arXiv},
       eprint = {1201.3359},
 primaryClass = {astro-ph.CO},
       adsurl = {https://ui.adsabs.harvard.edu/abs/2012MNRAS.424.1275B}
}

@ARTICLE{Spitzer1953,
       author = {{Spitzer}, Jr., Lyman and {Schwarzschild}, Martin},
        title = "{The Possible Influence of Interstellar Clouds on Stellar Velocities. II.}",
      journal = {\apj},
         year = 1953,
        month = jul,
       volume = {118},
        pages = {106},
          doi = {10.1086/145730},
       adsurl = {https://ui.adsabs.harvard.edu/abs/1953ApJ...118..106S}
}

@ARTICLE{Spitzer1951,
       author = {{Spitzer}, Jr., Lyman and {Schwarzschild}, Martin},
        title = "{The Possible Influence of Interstellar Clouds on Stellar Velocities.}",
      journal = {\apj},
         year = 1951,
        month = nov,
       volume = {114},
        pages = {385},
          doi = {10.1086/145478},
       adsurl = {https://ui.adsabs.harvard.edu/abs/1951ApJ...114..385S}
}

@ARTICLE{Wielen1977,
       author = {{Wielen}, R.},
        title = "{The Diffusion of Stellar Orbits Derived from the Observed Age-Dependence of the Velocity Dispersion}",
      journal = {\aap},
         year = 1977,
        month = sep,
       volume = {60},
       number = {2},
        pages = {263-275},
       adsurl = {https://ui.adsabs.harvard.edu/abs/1977A&A....60..263W}
}

@ARTICLE{Nordstrom2004,
       author = {{Nordstr{\"o}m}, B. and {Mayor}, M. and {Andersen}, J. and {Holmberg}, J. and {Pont}, F. and {J{\o}rgensen}, B.~R. and {Olsen}, E.~H. and {Udry}, S. and {Mowlavi}, N.},
        title = "{The Geneva-Copenhagen survey of the Solar neighbourhood. Ages, metallicities, and kinematic properties of {\ensuremath{\sim}}14 000 F and G dwarfs}",
      journal = {\aap},
         year = 2004,
        month = may,
       volume = {418},
        pages = {989-1019},
          doi = {10.1051/0004-6361:20035959},
archivePrefix = {arXiv},
       eprint = {astro-ph/0405198},
 primaryClass = {astro-ph},
       adsurl = {https://ui.adsabs.harvard.edu/abs/2004A&A...418..989N}
}

@ARTICLE{Minchev2006,
       author = {{Minchev}, I. and {Quillen}, A.~C.},
        title = "{Radial heating of a galactic disc by multiple spiral density waves}",
      journal = {\mnras},
         year = 2006,
        month = may,
       volume = {368},
       number = {2},
        pages = {623-636},
          doi = {10.1111/j.1365-2966.2006.10129.x},
archivePrefix = {arXiv},
       eprint = {astro-ph/0511037},
 primaryClass = {astro-ph},
       adsurl = {https://ui.adsabs.harvard.edu/abs/2006MNRAS.368..623M}
}

@ARTICLE{Bird2013,
       author = {{Bird}, Jonathan C. and {Kazantzidis}, Stelios and {Weinberg}, David H. and {Guedes}, Javiera and {Callegari}, Simone and {Mayer}, Lucio and {Madau}, Piero},
        title = "{Inside out and Upside down: Tracing the Assembly of a Simulated Disk Galaxy Using Mono-age Stellar Populations}",
      journal = {\apj},
         year = 2013,
        month = aug,
       volume = {773},
       number = {1},
          eid = {43},
        pages = {43},
          doi = {10.1088/0004-637X/773/1/43},
archivePrefix = {arXiv},
       eprint = {1301.0620},
 primaryClass = {astro-ph.GA},
       adsurl = {https://ui.adsabs.harvard.edu/abs/2013ApJ...773...43B}
}

@ARTICLE{Haywood2013,
       author = {{Haywood}, Misha and {Di Matteo}, Paola and {Lehnert}, Matthew D. and {Katz}, David and {G{\'o}mez}, Ana},
        title = "{The age structure of stellar populations in the solar vicinity. Clues of a two-phase formation history of the Milky Way disk}",
      journal = {\aap},
         year = 2013,
        month = dec,
       volume = {560},
          eid = {A109},
        pages = {A109},
          doi = {10.1051/0004-6361/201321397},
archivePrefix = {arXiv},
       eprint = {1305.4663},
 primaryClass = {astro-ph.GA},
       adsurl = {https://ui.adsabs.harvard.edu/abs/2013A&A...560A.109H}
}

@ARTICLE{Brook2004,
       author = {{Brook}, Chris B. and {Kawata}, Daisuke and {Gibson}, Brad K. and {Freeman}, Ken C.},
        title = "{The Emergence of the Thick Disk in a Cold Dark Matter Universe}",
      journal = {\apj},
         year = 2004,
        month = sep,
       volume = {612},
       number = {2},
        pages = {894-899},
          doi = {10.1086/422709},
archivePrefix = {arXiv},
       eprint = {astro-ph/0405306},
 primaryClass = {astro-ph},
       adsurl = {https://ui.adsabs.harvard.edu/abs/2004ApJ...612..894B}
}

@ARTICLE{Ma2017,
       author = {{Ma}, Xiangcheng and {Hopkins}, Philip F. and {Wetzel}, Andrew R. and {Kirby}, Evan N. and {Angl{\'e}s-Alc{\'a}zar}, Daniel and {Faucher-Gigu{\`e}re}, Claude-Andr{\'e} and {Kere{\v{s}}}, Du{\v{s}}an and {Quataert}, Eliot},
        title = "{The structure and dynamical evolution of the stellar disc of a simulated Milky Way-mass galaxy}",
      journal = {\mnras},
         year = 2017,
        month = may,
       volume = {467},
       number = {2},
        pages = {2430-2444},
          doi = {10.1093/mnras/stx273},
archivePrefix = {arXiv},
       eprint = {1608.04133},
 primaryClass = {astro-ph.GA},
       adsurl = {https://ui.adsabs.harvard.edu/abs/2017MNRAS.467.2430M}
}

@ARTICLE{Bovy2016,
       author = {{Bovy}, Jo and {Rix}, Hans-Walter and {Schlafly}, Edward F. and {Nidever}, David L. and {Holtzman}, Jon A. and {Shetrone}, Matthew and {Beers}, Timothy C.},
        title = "{The Stellar Population Structure of the Galactic Disk}",
      journal = {\apj},
         year = 2016,
        month = may,
       volume = {823},
       number = {1},
          eid = {30},
        pages = {30},
          doi = {10.3847/0004-637X/823/1/30},
archivePrefix = {arXiv},
       eprint = {1509.05796},
 primaryClass = {astro-ph.GA},
       adsurl = {https://ui.adsabs.harvard.edu/abs/2016ApJ...823...30B}
}

@ARTICLE{Minchev2017,
       author = {{Minchev}, I. and {Steinmetz}, M. and {Chiappini}, C. and {Martig}, M. and {Anders}, F. and {Matijevic}, G. and {de Jong}, R.~S.},
        title = "{The Relationship between Mono-abundance and Mono-age Stellar Populations in the Milky Way Disk}",
      journal = {\apj},
         year = 2017,
        month = jan,
       volume = {834},
       number = {1},
          eid = {27},
        pages = {27},
          doi = {10.3847/1538-4357/834/1/27},
archivePrefix = {arXiv},
       eprint = {1608.04737},
 primaryClass = {astro-ph.GA},
       adsurl = {https://ui.adsabs.harvard.edu/abs/2017ApJ...834...27M}
}

@ARTICLE{Juric2008,
       author = {{Juri{\'c}}, Mario and {Ivezi{\'c}}, {\v{Z}}eljko and {Brooks}, Alyson and {Lupton}, Robert H. and {Schlegel}, David and {Finkbeiner}, Douglas and {Padmanabhan}, Nikhil and {Bond}, Nicholas and {Sesar}, Branimir and {Rockosi}, Constance M. and {Knapp}, Gillian R. and {Gunn}, James E. and {Sumi}, Takahiro and {Schneider}, Donald P. and {Barentine}, J.~C. and {Brewington}, Howard J. and {Brinkmann}, J. and {Fukugita}, Masataka and {Harvanek}, Michael and {Kleinman}, S.~J. and {Krzesinski}, Jurek and {Long}, Dan and {Neilsen}, Jr., Eric H. and {Nitta}, Atsuko and {Snedden}, Stephanie A. and {York}, Donald G.},
        title = "{The Milky Way Tomography with SDSS. I. Stellar Number Density Distribution}",
      journal = {\apj},
         year = 2008,
        month = feb,
       volume = {673},
       number = {2},
        pages = {864-914},
          doi = {10.1086/523619},
archivePrefix = {arXiv},
       eprint = {astro-ph/0510520},
 primaryClass = {astro-ph},
       adsurl = {https://ui.adsabs.harvard.edu/abs/2008ApJ...673..864J}
}

@ARTICLE{FIRE2,
       author = {{Hopkins}, Philip F. and {Wetzel}, Andrew and {Kere{\v{s}}}, Du{\v{s}}an and {Faucher-Gigu{\`e}re}, Claude-Andr{\'e} and {Quataert}, Eliot and {Boylan-Kolchin}, Michael and {Murray}, Norman and {Hayward}, Christopher C. and {Garrison-Kimmel}, Shea and {Hummels}, Cameron and {Feldmann}, Robert and {Torrey}, Paul and {Ma}, Xiangcheng and {Angl{\'e}s-Alc{\'a}zar}, Daniel and {Su}, Kung-Yi and {Orr}, Matthew and {Schmitz}, Denise and {Escala}, Ivanna and {Sanderson}, Robyn and {Grudi{\'c}}, Michael Y. and {Hafen}, Zachary and {Kim}, Ji-Hoon and {Fitts}, Alex and {Bullock}, James S. and {Wheeler}, Coral and {Chan}, T.~K. and {Elbert}, Oliver D. and {Narayanan}, Desika},
        title = "{FIRE-2 simulations: physics versus numerics in galaxy formation}",
      journal = {\mnras},
         year = 2018,
        month = oct,
       volume = {480},
       number = {1},
        pages = {800-863},
          doi = {10.1093/mnras/sty1690},
archivePrefix = {arXiv},
       eprint = {1702.06148},
 primaryClass = {astro-ph.GA},
       adsurl = {https://ui.adsabs.harvard.edu/abs/2018MNRAS.480..800H}
}

@ARTICLE{Gurvich23,
       author = {{Gurvich}, Alexander B. and {Stern}, Jonathan and {Faucher-Gigu{\`e}re}, Claude-Andr{\'e} and {Hopkins}, Philip F. and {Wetzel}, Andrew and {Moreno}, Jorge and {Hayward}, Christopher C. and {Richings}, Alexander J. and {Hafen}, Zachary},
        title = "{Rapid disc settling and the transition from bursty to steady star formation in Milky Way-mass galaxies}",
      journal = {\mnras},
         year = 2023,
        month = feb,
       volume = {519},
       number = {2},
        pages = {2598-2614},
          doi = {10.1093/mnras/stac3712},
archivePrefix = {arXiv},
       eprint = {2203.04321},
 primaryClass = {astro-ph.GA},
       adsurl = {https://ui.adsabs.harvard.edu/abs/2023MNRAS.519.2598G}
}

@ARTICLE{Belokurov2022,
       author = {{Belokurov}, Vasily and {Kravtsov}, Andrey},
        title = "{From dawn till disc: Milky Way's turbulent youth revealed by the APOGEE+Gaia data}",
      journal = {\mnras},
         year = 2022,
        month = jul,
       volume = {514},
       number = {1},
        pages = {689-714},
          doi = {10.1093/mnras/stac1267},
archivePrefix = {arXiv},
       eprint = {2203.04980},
 primaryClass = {astro-ph.GA},
       adsurl = {https://ui.adsabs.harvard.edu/abs/2022MNRAS.514..689B}
}

@ARTICLE{Xiang22,
       author = {{Xiang}, Maosheng and {Rix}, Hans-Walter},
        title = "{A time-resolved picture of our Milky Way's early formation history}",
      journal = {\nat},
         year = 2022,
        month = mar,
       volume = {603},
       number = {7902},
        pages = {599-603},
          doi = {10.1038/s41586-022-04496-5},
archivePrefix = {arXiv},
       eprint = {2203.12110},
 primaryClass = {astro-ph.GA},
       adsurl = {https://ui.adsabs.harvard.edu/abs/2022Natur.603..599X}
}

@ARTICLE{Grand2016,
       author = {{Grand}, Robert J.~J. and {Springel}, Volker and {G{\'o}mez}, Facundo A. and {Marinacci}, Federico and {Pakmor}, R{\"u}diger and {Campbell}, David J.~R. and {Jenkins}, Adrian},
        title = "{Vertical disc heating in Milky Way-sized galaxies in a cosmological context}",
      journal = {\mnras},
         year = 2016,
        month = jun,
       volume = {459},
       number = {1},
        pages = {199-219},
          doi = {10.1093/mnras/stw601},
archivePrefix = {arXiv},
       eprint = {1512.02219},
 primaryClass = {astro-ph.GA},
       adsurl = {https://ui.adsabs.harvard.edu/abs/2016MNRAS.459..199G}
}

@ARTICLE{ruiz-lara2022,
       author = {{Ruiz-Lara}, T. and {Matsuno}, T. and {L{\"o}vdal}, S.~S. and {Helmi}, A. and {Dodd}, E. and {Koppelman}, H.~H.},
        title = "{Substructure in the stellar halo near the Sun. II. Characterisation of independent structures}",
      journal = {\aap},
         year = 2022,
        month = sep,
       volume = {665},
          eid = {A58},
        pages = {A58},
          doi = {10.1051/0004-6361/202243061},
archivePrefix = {arXiv},
       eprint = {2201.02405},
 primaryClass = {astro-ph.GA},
       adsurl = {https://ui.adsabs.harvard.edu/abs/2022A&A...665A..58R}
}

@ARTICLE{Navarro2018,
       author = {{Navarro}, Julio F. and {Yozin}, Cameron and {Loewen}, Nic and {Ben{\'\i}tez-Llambay}, Alejandro and {Fattahi}, Azadeh and {Frenk}, Carlos S. and {Oman}, Kyle A. and {Schaye}, Joop and {Theuns}, Tom},
        title = "{The innate origin of radial and vertical gradients in a simulated galaxy disc}",
      journal = {\mnras},
         year = 2018,
        month = may,
       volume = {476},
       number = {3},
        pages = {3648-3660},
          doi = {10.1093/mnras/sty497},
archivePrefix = {arXiv},
       eprint = {1709.01040},
 primaryClass = {astro-ph.GA},
       adsurl = {https://ui.adsabs.harvard.edu/abs/2018MNRAS.476.3648N}
}

@ARTICLE{gas2005,
       author = {{Kere{\v{s}}}, Du{\v{s}}an and {Katz}, Neal and {Weinberg}, David H. and {Dav{\'e}}, Romeel},
        title = "{How do galaxies get their gas?}",
      journal = {\mnras},
         year = 2005,
        month = oct,
       volume = {363},
       number = {1},
        pages = {2-28},
          doi = {10.1111/j.1365-2966.2005.09451.x},
archivePrefix = {arXiv},
       eprint = {astro-ph/0407095},
 primaryClass = {astro-ph},
       adsurl = {https://ui.adsabs.harvard.edu/abs/2005MNRAS.363....2K}
}

@ARTICLE{miglio2013,
       author = {{Miglio}, A. and {Chiappini}, C. and {Morel}, T. and {Barbieri}, M. and {Chaplin}, W.~J. and {Girardi}, L. and {Montalb{\'a}n}, J. and {Valentini}, M. and {Mosser}, B. and {Baudin}, F. and {Casagrande}, L. and {Fossati}, L. and {Silva Aguirre}, V. and {Baglin}, A.},
        title = "{Galactic archaeology: mapping and dating stellar populations with asteroseismology of red-giant stars}",
      journal = {\mnras},
         year = 2013,
        month = feb,
       volume = {429},
       number = {1},
        pages = {423-428},
          doi = {10.1093/mnras/sts345},
archivePrefix = {arXiv},
       eprint = {1211.0146},
 primaryClass = {astro-ph.GA},
       adsurl = {https://ui.adsabs.harvard.edu/abs/2013MNRAS.429..423M}
}

@ARTICLE{casagrande2016,
       author = {{Casagrande}, L. and {Silva Aguirre}, V. and {Schlesinger}, K.~J. and {Stello}, D. and {Huber}, D. and {Serenelli}, A.~M. and {Sch{\"o}nrich}, R. and {Cassisi}, S. and {Pietrinferni}, A. and {Hodgkin}, S. and {Milone}, A.~P. and {Feltzing}, S. and {Asplund}, M.},
        title = "{Measuring the vertical age structure of the Galactic disc using asteroseismology and SAGA}",
      journal = {\mnras},
         year = 2016,
        month = jan,
       volume = {455},
       number = {1},
        pages = {987-1007},
          doi = {10.1093/mnras/stv2320},
archivePrefix = {arXiv},
       eprint = {1510.01376},
 primaryClass = {astro-ph.GA},
       adsurl = {https://ui.adsabs.harvard.edu/abs/2016MNRAS.455..987C}
}

@ARTICLE{Anders2023,
       author = {{Anders}, F. and {Gispert}, P. and {Ratcliffe}, B. and {Chiappini}, C. and {Minchev}, I. and {Nepal}, S. and {Queiroz}, A.~B.~A. and {Amarante}, J.~A.~S. and {Antoja}, T. and {Casali}, G. and {Casamiquela}, L. and {Khalatyan}, A. and {Miglio}, A. and {Perottoni}, H. and {Schultheis}, M.},
        title = "{Spectroscopic age estimates for APOGEE red-giant stars: Precise spatial and kinematic trends with age in the Galactic disc}",
      journal = {\aap},
         year = 2023,
        month = oct,
       volume = {678},
          eid = {A158},
        pages = {A158},
          doi = {10.1051/0004-6361/202346666},
archivePrefix = {arXiv},
       eprint = {2304.08276},
 primaryClass = {astro-ph.GA},
       adsurl = {https://ui.adsabs.harvard.edu/abs/2023A&A...678A.158A}
}

@ARTICLE{Queiroz2023,
       author = {{Queiroz}, A.~B.~A. and {Anders}, F. and {Chiappini}, C. and {Khalatyan}, A. and {Santiago}, B.~X. and {Nepal}, S. and {Steinmetz}, M. and {Gallart}, C. and {Valentini}, M. and {Dal Ponte}, M. and {Barbuy}, B. and {P{\'e}rez-Villegas}, A. and {Masseron}, T. and {Fern{\'a}ndez-Trincado}, J.~G. and {Khoperskov}, S. and {Minchev}, I. and {Fern{\'a}ndez-Alvar}, E. and {Lane}, R.~R. and {Nitschelm}, C.},
        title = "{StarHorse results for spectroscopic surveys and Gaia DR3: Chrono-chemical populations in the solar vicinity, the genuine thick disk, and young alpha-rich stars}",
      journal = {\aap},
         year = 2023,
        month = may,
       volume = {673},
          eid = {A155},
        pages = {A155},
          doi = {10.1051/0004-6361/202245399},
archivePrefix = {arXiv},
       eprint = {2303.09926},
 primaryClass = {astro-ph.GA},
       adsurl = {https://ui.adsabs.harvard.edu/abs/2023A&A...673A.155Q}
}

@ARTICLE{khana2025,
       author = {{Khanna}, Shourya and {Yu}, Jie and {Drimmel}, Ronald and {Poggio}, Eloisa and {Cantat-Gaudin}, Tristan and {Castro-Ginard}, Alfred and {Kurbatov}, Evgeny and {Belokurov}, Vasily and {Brown}, Anthony and {Fouesneau}, Morgan and {Casey}, Andrew and {Rix}, Hans-Walter},
        title = "{GaiaUnlimited: The old stellar disc of the Milky Way as traced by the red clump}",
      journal = {\aap},
         year = 2025,
        month = sep,
       volume = {701},
          eid = {A270},
        pages = {A270},
          doi = {10.1051/0004-6361/202452798},
archivePrefix = {arXiv},
       eprint = {2410.22036},
 primaryClass = {astro-ph.GA},
       adsurl = {https://ui.adsabs.harvard.edu/abs/2025A&A...701A.270K}
}

@ARTICLE{Xiang2025,
       author = {{Xiang}, Maosheng and {Rix}, Hans-Walter and {Yang}, Hang and {Liu}, Jifeng and {Huang}, Yang and {Frankel}, Neige},
        title = "{The formation and survival of the Milky Way's oldest stellar disk}",
      journal = {Nature Astronomy},
         year = 2025,
        month = jan,
       volume = {9},
        pages = {101-110},
          doi = {10.1038/s41550-024-02382-w},
archivePrefix = {arXiv},
       eprint = {2410.09705},
 primaryClass = {astro-ph.GA},
       adsurl = {https://ui.adsabs.harvard.edu/abs/2025NatAs...9..101X}
}

@ARTICLE{Bensby2003,
       author = {{Bensby}, T. and {Feltzing}, S. and {Lundstr{\"o}m}, I.},
        title = "{Elemental abundance trends in the Galactic thin and thick disks as traced by nearby F and G dwarf stars}",
      journal = {\aap},
         year = 2003,
        month = nov,
       volume = {410},
        pages = {527-551},
          doi = {10.1051/0004-6361:20031213},
       adsurl = {https://ui.adsabs.harvard.edu/abs/2003A&A...410..527B}
}

@ARTICLE{Thulasidharan2026,
       author = {{Thulasidharan}, Lekshmi and {D'Onghia}, Elena and {Benjamin}, Robert and {Drimmel}, Ronald and {Poggio}, Eloisa},
        title = "{The Age-Thickness Relation as a Tracer of the Merger History of Disk Galaxies}",
      journal = {arXiv e-prints},
         year = 2026,
        month = jul,
          eid = {arXiv:2607.19537},
        pages = {arXiv:2607.19537},
archivePrefix = {arXiv},
       eprint = {2607.19537},
 primaryClass = {astro-ph.GA},
       adsurl = {https://ui.adsabs.harvard.edu/abs/2026arXiv260719537T}
}

@ARTICLE{Feltzing2026,
       author = {{Feltzing}, Sofia and {Feuillet}, Diane and {Bensby}, Thomas},
        title = "{Dawn of the Milky Way Disk: Determination of When a Rotationally Supported Disk Appears and Dating the Spin-up of the Disk}",
      journal = {\apjl},
         year = 2026,
        month = jun,
       volume = {1004},
       number = {2},
          eid = {L28},
        pages = {L28},
          doi = {10.3847/2041-8213/ae6f14},
archivePrefix = {arXiv},
       eprint = {2605.20959},
 primaryClass = {astro-ph.GA},
       adsurl = {https://ui.adsabs.harvard.edu/abs/2026ApJ..1004L..28F}
}

@ARTICLE{Fuhrmann1998,
       author = {{Fuhrmann}, Klaus},
        title = "{Nearby stars of the Galactic disk and halo}",
      journal = {\aap},
         year = 1998,
        month = oct,
       volume = {338},
        pages = {161-183},
       adsurl = {https://ui.adsabs.harvard.edu/abs/1998A&A...338..161F}
}

@ARTICLE{Bensby2014,
       author = {{Bensby}, T. and {Feltzing}, S. and {Oey}, M.~S.},
        title = "{Exploring the Milky Way stellar disk. A detailed elemental abundance study of 714 F and G dwarf stars in the solar neighbourhood}",
      journal = {\aap},
         year = 2014,
        month = feb,
       volume = {562},
          eid = {A71},
        pages = {A71},
          doi = {10.1051/0004-6361/201322631},
archivePrefix = {arXiv},
       eprint = {1309.2631},
 primaryClass = {astro-ph.GA},
       adsurl = {https://ui.adsabs.harvard.edu/abs/2014A&A...562A..71B}
}

@ARTICLE{Reddy2003,
       author = {{Reddy}, Bacham E. and {Tomkin}, Jocelyn and {Lambert}, David L. and {Allende Prieto}, Carlos},
        title = "{The chemical compositions of Galactic disc F and G dwarfs}",
      journal = {\mnras},
         year = 2003,
        month = mar,
       volume = {340},
       number = {1},
        pages = {304-340},
          doi = {10.1046/j.1365-8711.2003.06305.x},
archivePrefix = {arXiv},
       eprint = {astro-ph/0211551},
 primaryClass = {astro-ph},
       adsurl = {https://ui.adsabs.harvard.edu/abs/2003MNRAS.340..304R}
}

@ARTICLE{Adybekyan2011,
       author = {{Adibekyan}, V. Zh. and {Santos}, N.~C. and {Sousa}, S.~G. and {Israelian}, G.},
        title = "{A new {\ensuremath{\alpha}}-enhanced super-solar metallicity population}",
      journal = {\aap},
         year = 2011,
        month = nov,
       volume = {535},
          eid = {L11},
        pages = {L11},
          doi = {10.1051/0004-6361/201118240},
archivePrefix = {arXiv},
       eprint = {1111.4936},
 primaryClass = {astro-ph.GA},
       adsurl = {https://ui.adsabs.harvard.edu/abs/2011A&A...535L..11A}
}

@ARTICLE{Bensby2004,
       author = {{Bensby}, T. and {Feltzing}, S. and {Lundstr{\"o}m}, I.},
        title = "{A possible age-metallicity relation in the Galactic thick disk?}",
      journal = {\aap},
         year = 2004,
        month = jul,
       volume = {421},
        pages = {969-976},
          doi = {10.1051/0004-6361:20035957},
archivePrefix = {arXiv},
       eprint = {astro-ph/0403591},
 primaryClass = {astro-ph},
       adsurl = {https://ui.adsabs.harvard.edu/abs/2004A&A...421..969B}
}

@ARTICLE{Bensby2007,
       author = {{Bensby}, T. and {Zenn}, A.~R. and {Oey}, M.~S. and {Feltzing}, S.},
        title = "{Tracing the Galactic Thick Disk to Solar Metallicities}",
      journal = {\apjl},
         year = 2007,
        month = jul,
       volume = {663},
       number = {1},
        pages = {L13-L16},
          doi = {10.1086/519792},
archivePrefix = {arXiv},
       eprint = {0705.2060},
 primaryClass = {astro-ph},
       adsurl = {https://ui.adsabs.harvard.edu/abs/2007ApJ...663L..13B}
}

% %%%%%%%%%%%%%%%%%%%%%%%%%%%%%%%%%%%%%%%%%%%%%%%%%%%%%%%%%%%%%%

\begin{appendix}

\section{Selection of \texttt{cold} and \texttt{hot} subsets}\label{appendix:selection}

The selection was carried out by assigning each star a probability of belonging to the \texttt{hot}, \texttt{cold}, or \texttt{halo} subset, assuming a Gaussian distribution for the Galactic velocity components $(U, V, W)$. The probability $p_i$ of a star belonging to component $i$ is defined as:

\begin{align}
    w_i =& \,\rho_{0, i} \, \exp\left[-\tfrac{1}{2} \left( 
    \left(\frac{U - \mu_{U, i}}{\sigma_{U, i}}\right)^2 + 
    \left(\frac{V - \mu_{V, i}}{\sigma_{V, i}}\right)^2 + \right.\right. \nonumber\\ 
    & + \left.\left. \left(\frac{W - \mu_{W, i}}{\sigma_{W, i}}\right)^2 \right)\right] \\
    p_i =& \,\frac{w_i}{\sum\limits_j w_j}
\end{align}
where $\rho_{0,i}$ is the local density normalization of population $i$, $\mu_{U, i}$, $\mu_{V, i}$, and $\mu_{W, i}$ are the assumed mean velocities, and $\sigma_{U, i}$, $\sigma_{V, i}$, and $\sigma_{W, i}$ are the corresponding assumed velocity dispersions of that component. For this work, we select the stars with $p_i>0.75$, so no overlap is possible between subsets. The same parameters as those in \citetalias{emma2025} are used \citep{Chiba2000, Soubiran2003, Ramirez2013, Reid2019}. The specific values are presented in Table~A1 in \citetalias{Gallart2024}. 

Our dynamical selection of the \texttt{hot} subset naturally filters out halo stars. As an additional check, we determined the deSFH of the halo within the same volume. The range of ages is similar in the \texttt{hot} and \texttt{halo} subsets, but the AMD of both is clearly distinct. In any case, \texttt{halo} stars represent less than 10\% of the total \texttt{hot} subset. Therefore, any actual halo contamination must be well below this $10\%$ upper limit. We do not expect halo contamination to impact our results in this work.

\section{Example Solutions}\label{appendixCMDfit}

\begin{figure*}
   \centering
   \includegraphics[width=\textwidth]{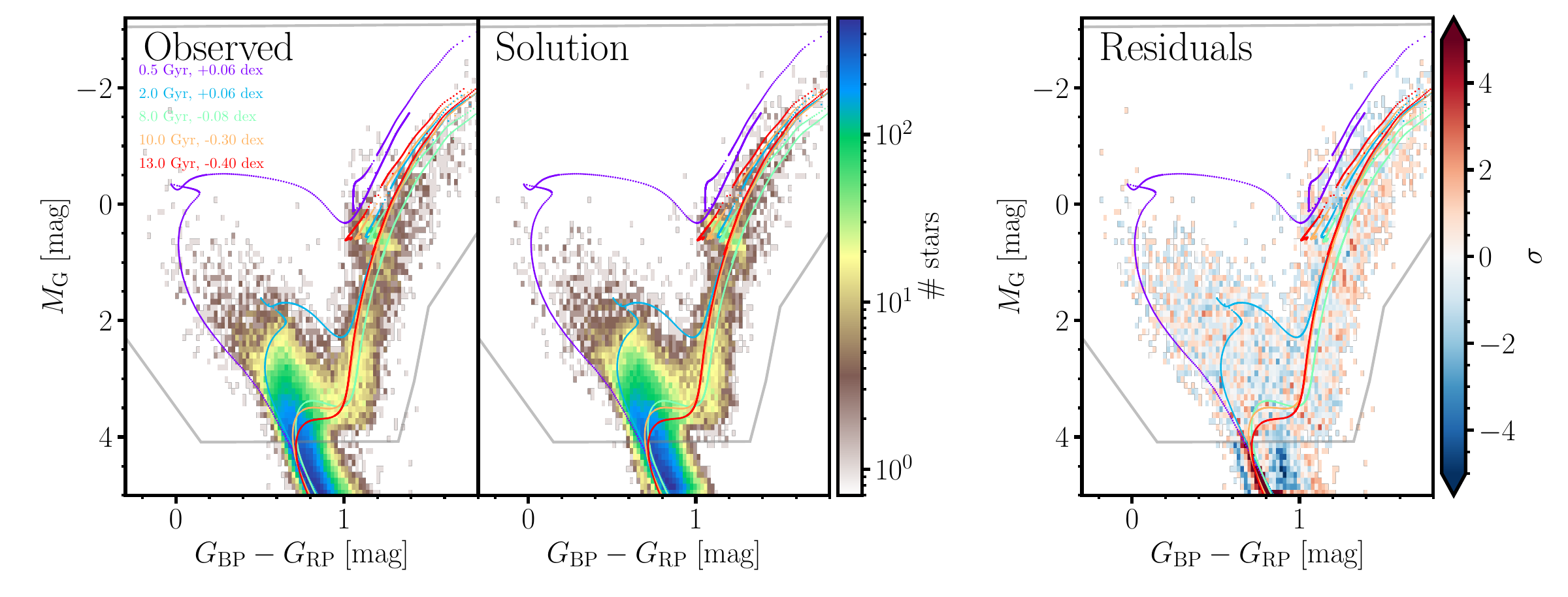}
      \caption{From left to right: Observed CMD, solution CMD, and residuals of the fourth layer ($0.3\;\text{kpc}<|Z|<0.4\;\text{kpc}$) of the \texttt{cold} subset. Overlaid on all panels are 5 BaSTI-IAC isochrones for reference, with ages and metallicities indicated by colors in the left panel. The gray polygon represents the bundle where we perform the CMD fitting.}
    \label{fig:cmdfitall}
\end{figure*}

Figure~\ref{fig:cmdfitall} displays, from left to right, the observed CMD, the solution CMD, and the residuals\footnote{Residuals are defined as~$\sigma~=~(\text{observed}~-~\text{solution})/\sqrt{\text{observed} + \text{solution}}$} for the fourth layer ($0.3\;\text{kpc}<|Z|<0.4\;\text{kpc}$) of the \texttt{cold} subset as an example of the fit quality. Overlaid on all panels are 5 BaSTI-IAC isochrones for reference, with their ages and metallicities indicated by colors in the left panel. The gray polygon represents the region used for CMD fitting. 

\begin{figure*}
   \centering
   \includegraphics[width=0.95\textwidth]{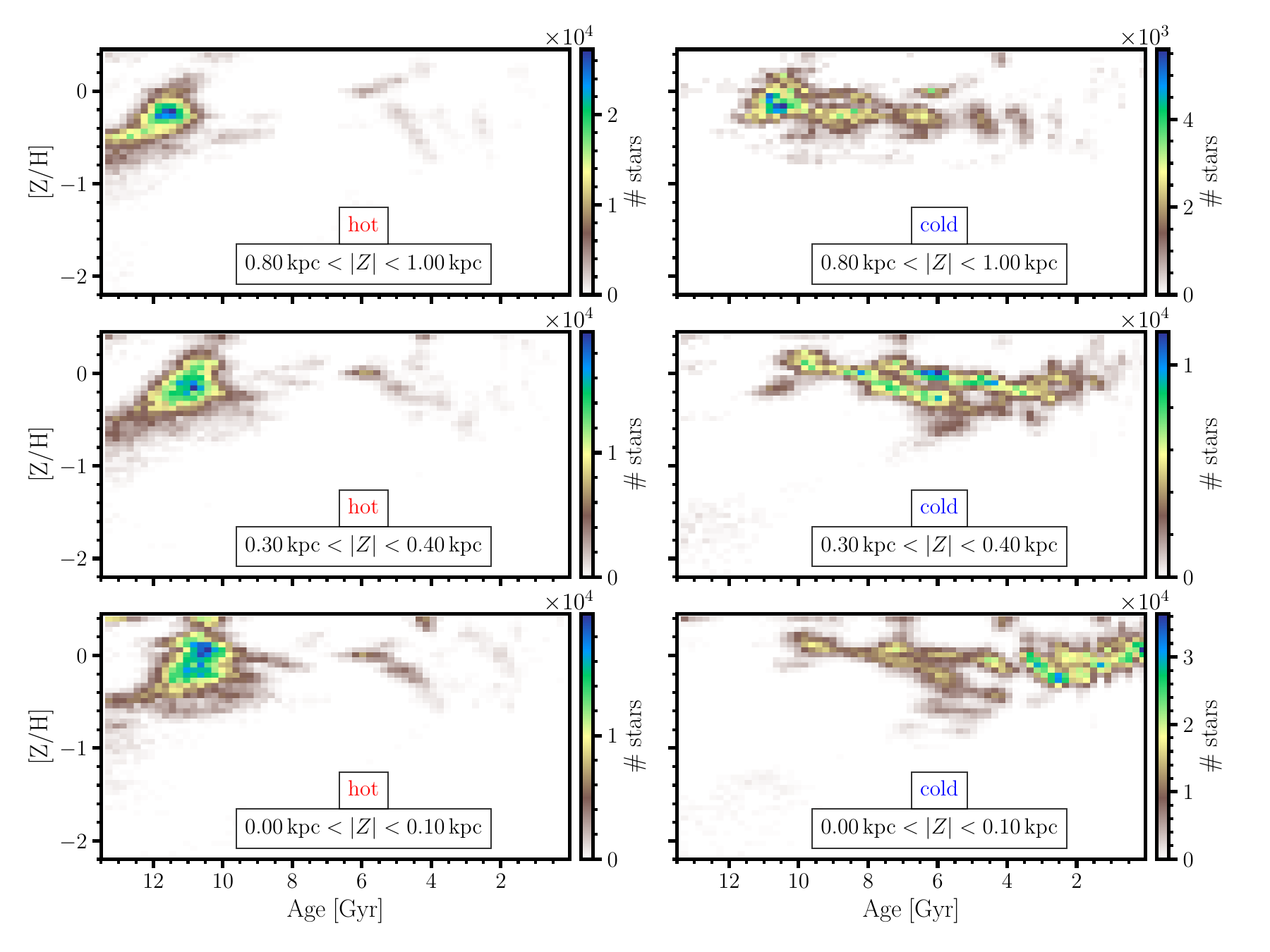}
      \caption{AMDs corrected for all completeness effects ($N_i(\text{Age, [Z/H]})$) displayed across 6 panels. Each column corresponds to a different subset: left column for the \texttt{hot} and right column for the \texttt{cold}. Each row represents a distinct layer, with distance from the galactic plane increasing from bottom to top.}
         \label{fig:solutions_all}
\end{figure*}

Figure~\ref{fig:solutions_all} shows a selection of three solutions for each subset corrected for all completeness effects (i.e. $N_i(\text{Age, [Z/H]})$). The left column corresponds to the \texttt{hot} and the right column to the \texttt{cold} subset. Each row represents a different layer of the sample, moving progressively farther from the Galactic plane from the lower to the upper panel.

As already observed in \citetalias{emma2025}, a significant difference, particularly in age, is evident here between the \texttt{hot} and \texttt{cold} subsets. Note that the AMDs of the different layers of the \texttt{hot} subset are very similar, indicating great homogeneity among stellar populations of the \texttt{hot} disk in the analysed volume. In contrast, in the \texttt{cold} subset, young populations become progressively more concentrated toward the plane.
    
\section{Dependence on the age metallicity binning scheme}\label{appendixqMAMs}

\begin{figure}
    \centering
    \includegraphics[width=\linewidth]{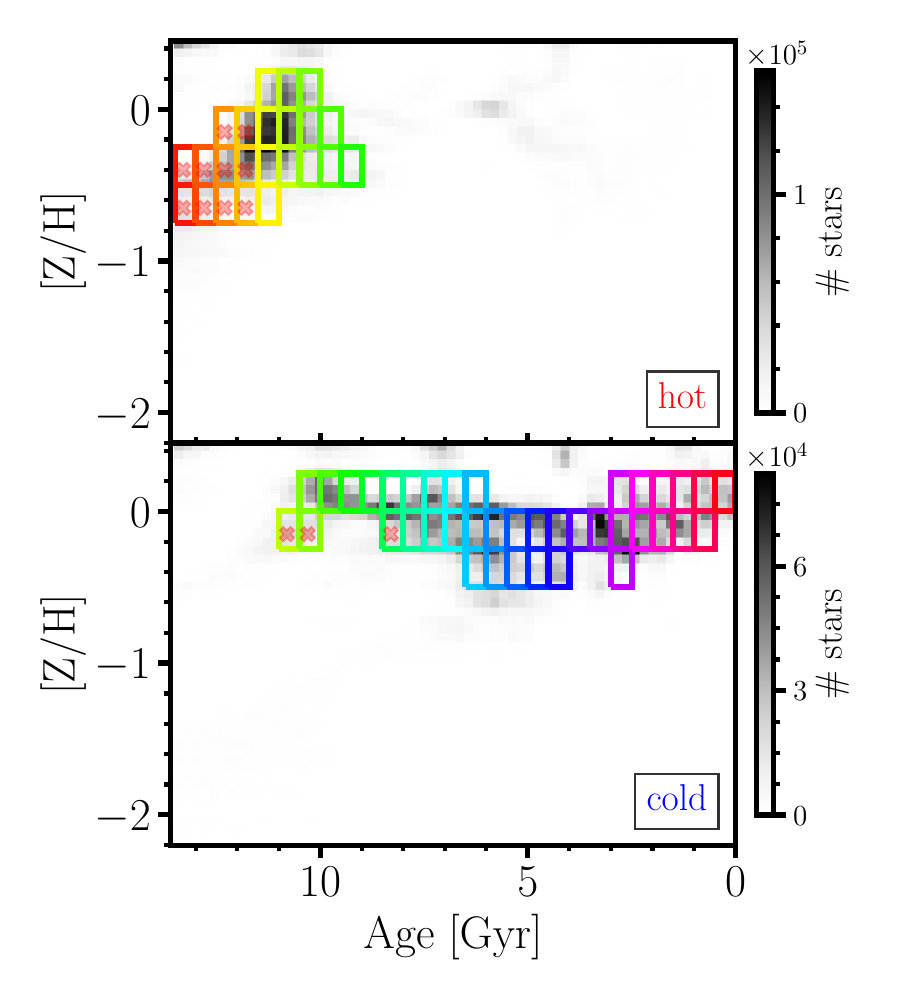}
    \caption{Age-metallicity distributions of the \texttt{hot} (top panel) and \texttt{cold} (bottom panel) samples. The overlaid uniform grid shows the bins that satisfy the requirement of containing more than 1\% of the stars in the corresponding sample. Red crosses mark bins for which the derived scale height has an uncertainty larger than 30\%.}
    \label{fig:grid_age_metal}
\end{figure}

The qMAMs were defined using the age–metallicity division described in Section~\ref{methodology:qmams}. Since the definition of these populations could potentially influence the measured $h_Z$-age relation, we performed an additional test using a uniform age–metallicity grid. We divided the age-metallicity plane into bins of $0.5$ Gyr in age and $0.25$ dex in metallicity, covering ranges of $0-13.5$ Gyr and $-1.25 \leq \text{[Z/H]} \leq 0.25$ dex. Figure~\ref{fig:grid_age_metal} shows the age-metallicity distributions of the hot and cold samples, together with the resulting grid after applying the requirement that each bin should contain more than 1\% of the stars in the corresponding sample. The top panel corresponds to the \texttt{hot} sample, while the bottom panel shows the \texttt{cold} sample. Bins marked with red crosses indicate scale-height measurements with uncertainties larger than 30\%.

\begin{figure}
    \centering
    \includegraphics[width=\linewidth]{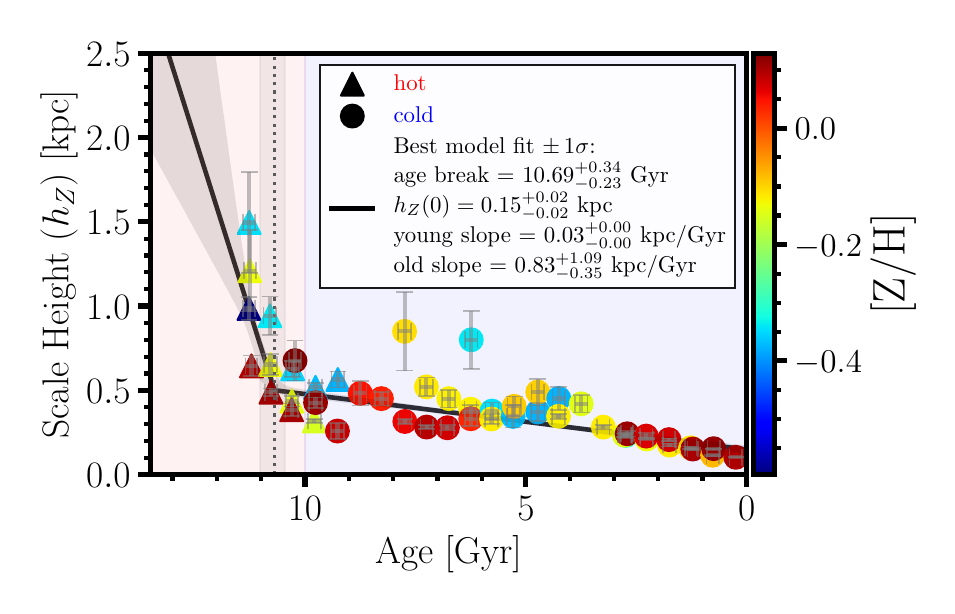}
    \caption{Same as Figure~\ref{fig:scale_height} but using the selected bins in the uniform grid. The line ($\pm 1\sigma$) and the parameters in the legend represent a MCMC fit for this data.}
    \label{fig:age_hz_grid}
\end{figure}

For each selected bin, we measured the vertical scale height following exactly the same procedure adopted in the main analysis (see Section~\ref{section:methodology}). The resulting $h_Z$-age relations are shown in Figure~\ref{fig:age_hz_grid}. The overall behaviour remains unchanged with this grid. The scale height increases with age in both age regimes, and the transition around $\sim10.7-10.8$ Gyr is recovered independently of the adopted binning scheme. The parameters obtained with this new grid are consistent within the uncertainties with those obtained using the original qMAMs. The largest difference is found in the slope of the relation at old ages, which is $1.88^{+1.38}_{-0.83}$ kpc/Gyr for the original qMAMs, compared to $0.74^{+0.48}_{-0.28}$ kpc/Gyr for the uniform grid. However, this still represents a large difference compared to the slope at young ages, as highlighted in the main body of the Letter. We did not adopt this uniform-grid approach in the Letter because we believe it is physically more meaningful to map the stellar populations that are clearly identifiable in the age–metallicity distributions, rather than applying uniform cuts that split several of these populations across multiple bins. Nevertheless, these results indicate that the two-stage settling of the Galactic disk reported in the main body is not driven by the particular construction of the qMAMs, but reflects a robust feature of the data.

\section{Modelling the scale-height vs age relation}\label{appendix:mcmc}
We model the relation between stellar age and scale height using a broken linear model. This model consists of two linear regimes separated by a transition age, which is treated as a free parameter:

\begin{equation}
    f(\text{age}) =
    \begin{cases}
        a + b_{\rm young}\cdot\text{age}, & \text{age} < \text{age}_\text{break}, \\[4pt]
        a + b_{\rm young}\cdot\text{age}_\text{break} + 
        & \text{age} \ge \text{age}_\text{break},\\+ \,b_{\rm old}\,(\text{age}-\text{age}_\text{break}), 
    \end{cases}
\end{equation}

where $a$ is the intercept, $b_{\rm young}$ and $b_{\rm old}$ are the slopes of the young and old age regimes, respectively, and $\text{age}_\text{break}$ is the transition age. 

We determine the posterior distribution of these parameters using an MCMC method. In addition to the model parameters, we include an intrinsic scatter, $\sigma_\text{int}$, to account for the dispersion in scale height that is not explained by the fitted relation or by the measurement uncertainties. Assuming Gaussian residuals, the total variance associated with each data point is
\begin{equation}
    \sigma_i^2 =
    \sigma_{\rm int}^2
    + \sigma_{h_Z,i}^2
    + \left(\frac{df}{d\text{age}}\sigma_{\text{age},i}\right)^2,
\end{equation}

where $\sigma_{h_Z,i}$ and $\sigma_{\text{age},i}$ are the uncertainties in scale height and age, respectively. The last term accounts for the contribution of the age uncertainty to the uncertainty in the predicted scale height. The intrinsic scatter is treated as a free parameter in the MCMC and is constrained to be positive. The likelihood is
\begin{equation}
\ln \mathcal{L} = -\frac{1}{2} \sum_i \left[\frac{(h_{Z,i}-f(\text{age}_i))^2}{\sigma_i^2} + \ln(2\pi\sigma_i^2) \right],
\end{equation}

where $h_{Z,i}$ and $\text{age}_i$ are the measured scale height and age of the qMAM $i$, respectively, and $f(\text{age}_i)$ is the scale height predicted by the model.

We use uniform priors for the model parameters. The slopes are restricted to $[-5,5]$ kpc/Gyr, while the transition age is allowed to vary over the full age range. The MCMC sampling is performed using 48 walkers and 16000 steps. An initial burn-in phase of 3000 steps is used to allow the walkers to converge towards the high-probability region before obtaining the final posterior distribution. 

\begin{figure*}
    \centering
    \includegraphics[width=\linewidth]{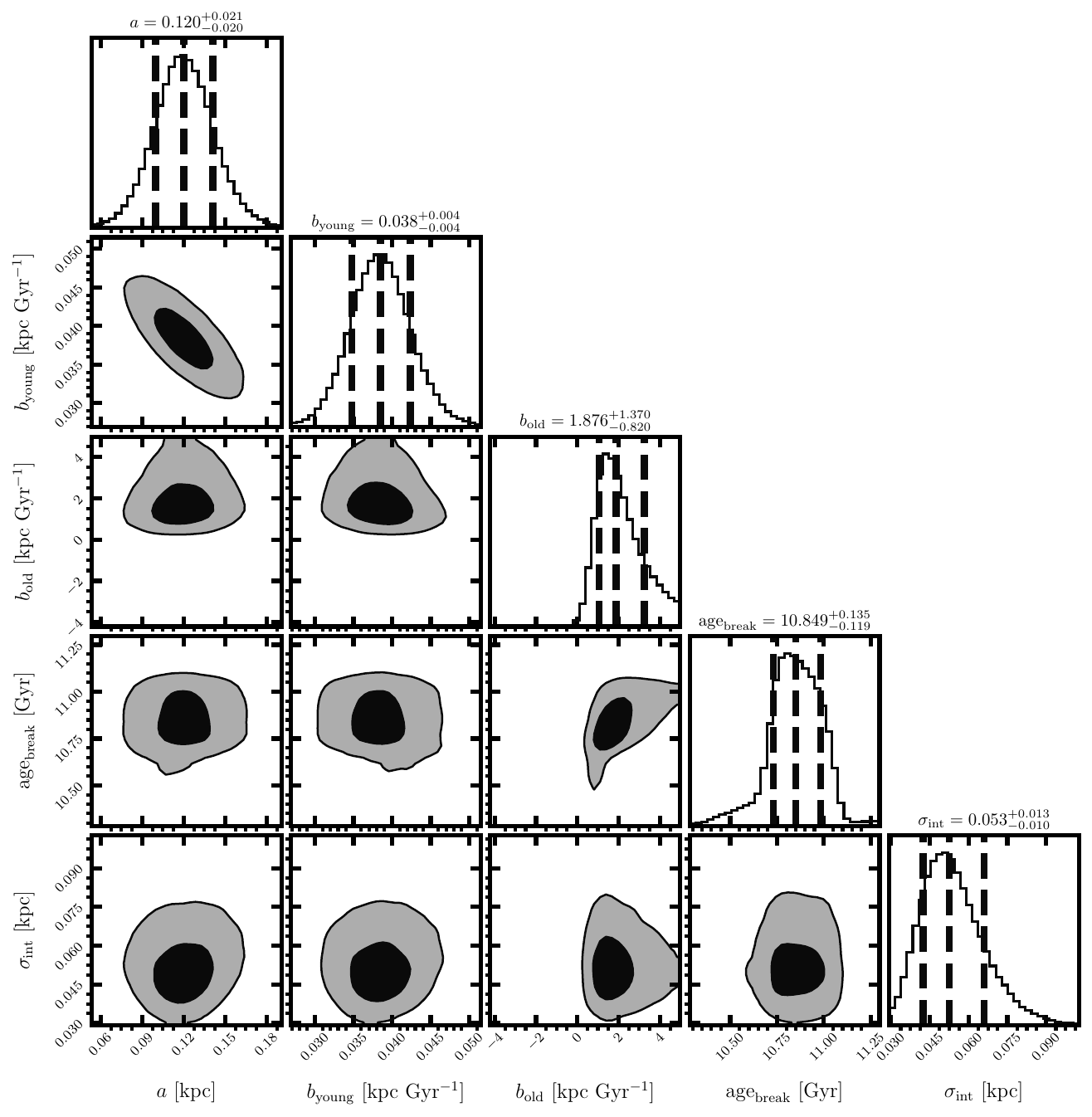}
    \caption{The posterior distributions for the intercept, the young-age slope, the old-age slope, the transition age, and the intrinsic scatter are shown for the fit to the qMAMs presented in Figure~\ref{fig:scale_height}. The contours indicate the 1 and 2 $\sigma$ confidence regions. The dashed vertical lines mark the 16th, 50th, and 84th percentiles of the posterior distributions.}
    \label{fig:corner_qmams}
\end{figure*}

Figure~\ref{fig:corner_qmams} shows the corner plot of the posterior distributions of the model parameters obtained from the data shown in Figure~\ref{fig:scale_height}. The distributions are well behaved and all parameters are constrained by the data. The intrinsic scatter found is negligible, which is why it was not included in the main discussion. No degeneracies are observed between the fitted parameters, indicating that the model is sufficiently constrained.

\section{The \texttt{hot} Disk qMAMs and the Special Case of \texttt{o2} and \texttt{o3}}\label{appendix:k2}

\begin{figure}[htbp]
   \centering
   \includegraphics[width=\columnwidth]{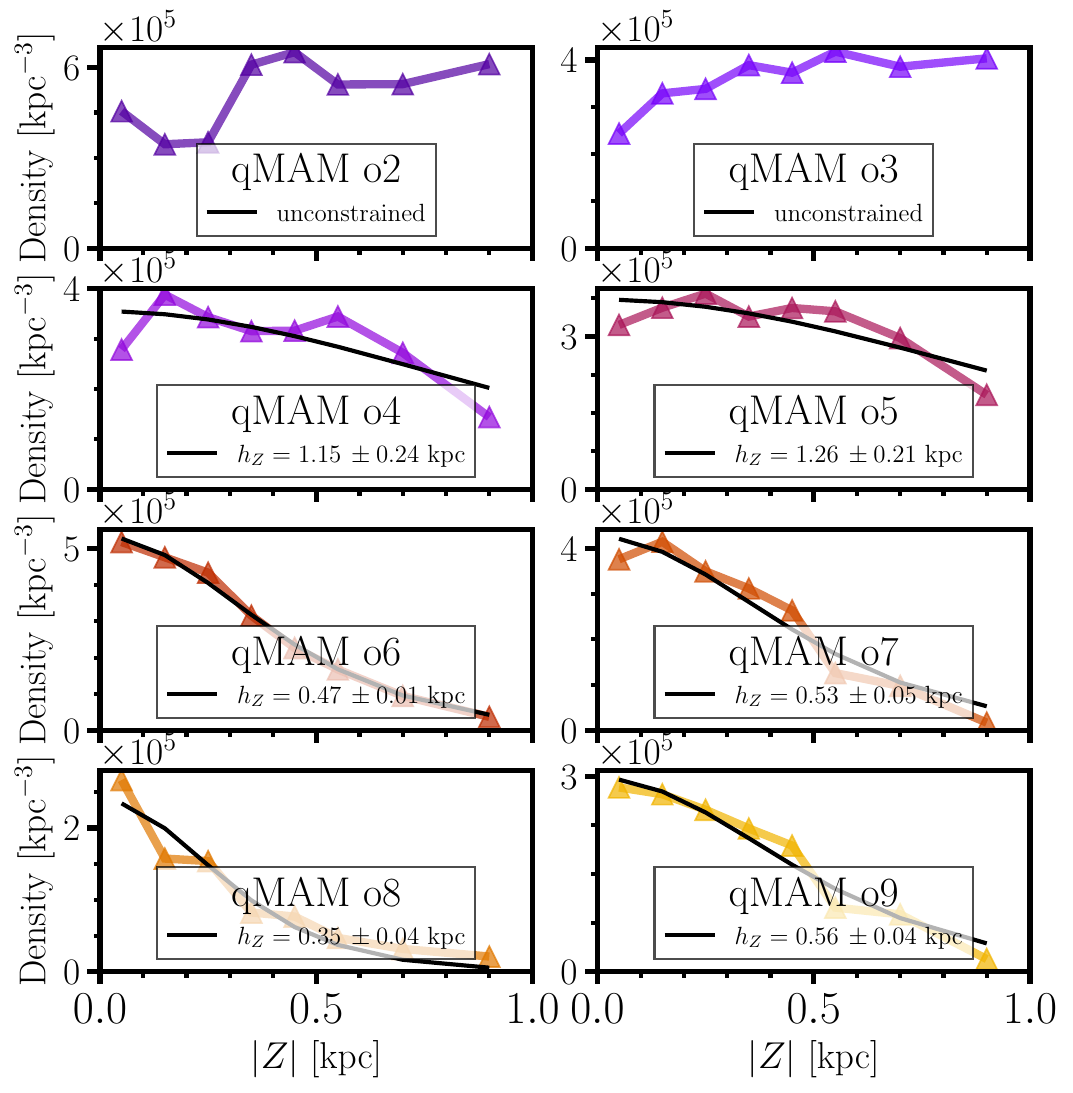}
      \caption{Vertical absolute density profiles $\rho(|Z|, \text{qMAM})$ for the main qMAMs identified within the kinematically selected \texttt{hot} component. Solid black lines represent the best-fit  using a $\operatorname{sech}^{2}$ function. The derived $h_Z$ and its associated $1\sigma$ standard error are specified in the legend of each panel.}
         \label{fig:profiles_thick_nok2}
\end{figure}

Figure~\ref{fig:profiles_thick_nok2} displays the vertical density profiles for the main qMAMs identified in the \texttt{hot} subset. The best fit scale height and its corresponding error are indicated in the legend of each panel. While most populations exhibit vertical profiles that are successfully reproduced by the fitted function, qMAM \texttt{o2} and \texttt{o3} remain basically flat within $|Z| <$1 kpc preventing an $h_Z$ fit.

To overcome this limitation and obtain an estimate of their scale height, we extended our analysis using an alternative volume selection. This new subset (called \texttt{disk}) has $R<250$ pc but spans up to $|Z|<3.5$ kpc. A full kinematic selection cannot be performed up to this distance, but halo stars were removed to the first order from the sample using a tangential velocity cut ($v_t<200$ km s$^{-1}$). Since these highly distinct qMAMs occupy a region of the age-metallicity plane completely devoid of \texttt{cold} disk stars, we can trace their profile from the selection of this subset. 
        
We have divided this tall sample with the following Z limits: 

\begin{align}
   |Z| \; \text{layers} =& [ 0.0 , 0.05 , 0.1 , 0.15 , 0.2 , 0.3 , 0.4 , 0.5 , 0.6 , 0.7 ,\nonumber\\ &0.8, 0.9 , 1.0 , 1.2 , 1.6 , 2.3 , 3.5 ]\; \text{kpc} \nonumber
\end{align}

\begin{figure}
   \centering
   \includegraphics[width=\columnwidth]{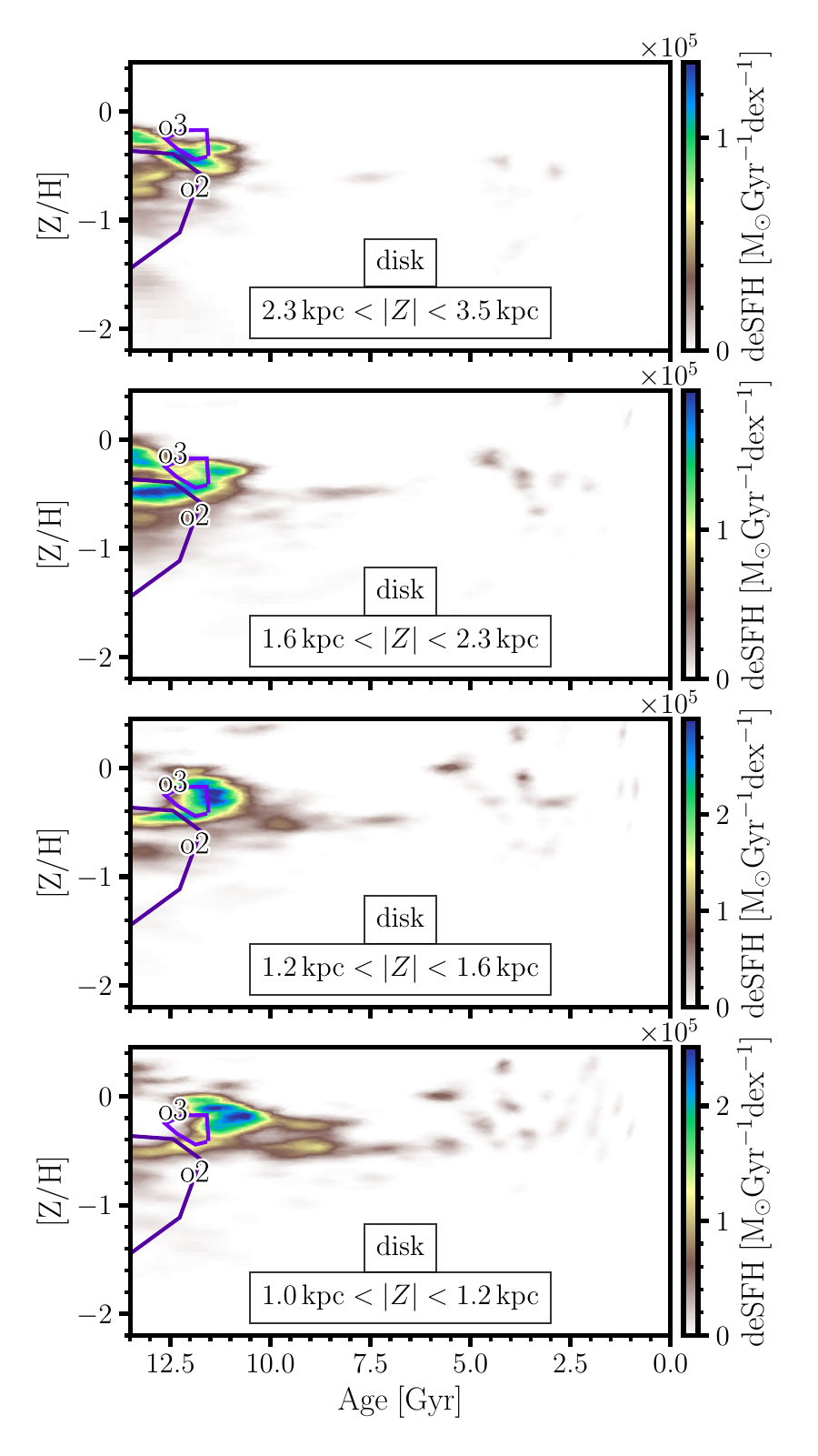}
      \caption{deSFH of four layers from the complete sample (without the halo). Each row represents a distinct layer, with distance from the Galactic plane increasing from bottom to top. qMAMs \texttt{o2} and \texttt{o3} are overlaid in continuos lines for reference.}
         \label{fig:amds_k2}
\end{figure}

Figure~\ref{fig:amds_k2} shows four deSFHs from different layers of this subset, ranging from 1 kpc to 3.5 kpc (from bottom to top in the panels). We present the deSFHs in this plot rather than the number of stars because at these vertical heights the number of stars is inherently low producing noisy plots. The deSFH highlights transitions between layers more clearly. As $|Z|$ increases, \texttt{o2} becomes predominant in the number of stars compared to the rest, being practically the only qMAM present between 2.3 and 3.5 kpc. It is easily distinguishable in the layers with $|Z|>$ 1 kpc where the \texttt{hot} disk dominates. Therefore, utilizing this larger vertical subset allows us to map the profile of \texttt{o2} and \texttt{o3} at higher $|Z|$ without being limited by the completeness of radial velocity measurements.
    
\begin{figure}
   \centering
   \includegraphics[width=\columnwidth]{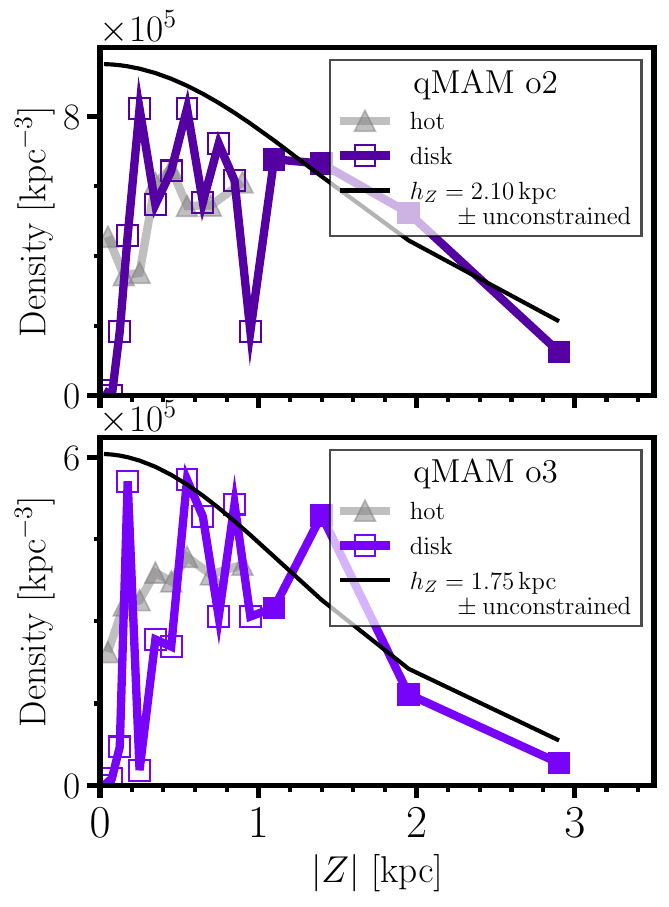}
      \caption{Extended vertical density profile for the qMAM \texttt{o2} in orange (top panel) and for the qMAM \texttt{o3} in green (bottom panel), derived using the geometrically selected sample ($R < 250$~pc, $|Z| \le 3.5$~kpc). The solid black lines indicate the formal fit restricted exclusively to the four outermost vertical layers ($|Z| > 1$~kpc), the filled squares. The gray triangles represent the same qMAMs in each panel but for the \texttt{hot} subset for reference.}
         \label{fig:profile_k2}
\end{figure}

The resulting extended vertical profile for qMAMs \texttt{o2} and \texttt{o3} (top and bottom, respectively) are presented in Figure~\ref{fig:profile_k2}. The distributions appear noticeably noisy, particularly near the Galactic plane. In these low-$Z$ layers, the CMD is heavily dominated by young populations, which complicates the accurate recovery of minor old and metal-poor components, which share position in the CMD with the more abundant younger, more metal-rich populations. In fact, up to $|Z|$=150 pc, these populations are not recovered at all, and in higher layers, they are recovered in lower numbers than expected, as we have verified using tests with mock stellar populations. In Figure~\ref{fig:profile_k2} we have also superimposed the \texttt{o2} and \texttt{o3} profile up to 1 kpc derived from the \texttt{hot} subset. This profile (in gray) shows a less noisy behaviour and demonstrates that the density derived for \texttt{o2} and \texttt{o3} in both samples is consistent.  

For the combined analysis of both spatial selections, we constrain a range of possible values for the $h_Z$ of \texttt{o2} and \texttt{o3} in the following way:
i) Since the kinematically selected sample shows no drop in density up to 1 kpc, and remains flatter than qMAM \texttt{o5} (see Figure~\ref{fig:profiles_thick_nok2}), their $h_Z$ must be larger than that of \texttt{o5}, that is, $h_Z>1.26$ kpc; ii) for the extended high-altitude layers we restrict our fit exclusively to the four highest vertical layers ($|Z|>1$ kpc) for which the density shows a less noisy behaviour and decays, thanks to the fact that the characterization of these old, relatively minor populations is less affected by the contamination by young stars in them. The fit to these four layers yields a scale height estimate of $h_Z=2.10$ kpc and $h_Z=1.75$ kpc for \texttt{o2} and \texttt{o3}, respectively. We do not consider that the errors obtained by the fitting method represent the true uncertainty we have regarding these populations. Therefore, we can not derive a secure upper limit. We decided to adopt an estimate of $h_Z=2.10$ kpc and $h_Z=1.75$ kpc with $h_Z>1.26$ kpc as their lower limit. This highly conservative estimate guarantees that our assessment about qMAMs \texttt{o2} and \texttt{o3} remains robust, strongly reinforcing the rapid early vertical settling and the upside-down evolutionary framework discussed in this Letter. 
        
\FloatBarrier

\end{appendix}
\end{document}